\documentclass[reprint,amsmath,amssymb,aps,prb]{revtex4-1}
\usepackage{graphicx}
\usepackage{xcolor}
\usepackage{dcolumn}
\usepackage{bm}
\usepackage{amsmath}
\usepackage[normalem]{ulem}
\usepackage{siunitx}
\usepackage{color,soul}
\usepackage[colorlinks=true, citecolor=red, urlcolor=blue]{hyperref}
\usepackage[para,online,flushleft]{threeparttable}
\makeatletter
\def\blfootnote{\xdef\@thefnmark{}\@footnotetext}
\makeatother

\begin{document}

\preprint{APS/123-QED}
\title{Ionic vs. van der Waals Layered Materials: \\
Identification and Comparison of Elastic Anisotropy}
\author{Robert McKinney$^{1,4}$, Prashun Gorai$^{2,4}$, Sukriti Manna$^{3,4}$, Eric Toberer$^{1,4}$, and Vladan Stevanovi\'{c}$^{2,4}$
\footnote{Corresponding author, email: vstevano@mines.edu}}
\affiliation{$^1$Dept. of Physics, Colorado School of Mines, Golden, Colorado 80401, USA\\
$^2$Dept. of Metallurgical and Materials Engineering, Colorado School of Mines, Golden, Colorado 80401, USA\\
$^3$Dept. of Mechanical Engineering, Colorado School of Mines, Golden, Colorado 80401, USA\\
$^4$National Renewable Energy Laboratory, Golden, CO 80401, USA}
\date{\today}

\begin{abstract}
In this work, we expand the set of known layered compounds to include ionic layered materials, which are well known for superconducting, thermoelectric, and battery applications. Focusing on known ternary compounds from the ICSD, we screen for ionic layered structures by expanding upon our previously developed algorithm for identification of van der Waals (vdW) layered structures, thus identifying over 1,500 ionic layered compounds. Since vdW layered structures can be chemically mutated to form ionic layered structures, we have developed a methodology to structurally link binary vdW to ternary ionic layered materials. We perform an in-depth analysis of similarities and differences between these two classes of layered systems and assess the interplay between layer geometry and bond strength with a study of the elastic anisotropy. We observe a rich variety of anisotropic behavior in which the layering direction alone is not a simple predictor of elastic anisotropy. Our results enable discovery of new layered materials through intercalation or de-intercalation of vdW or ionic layered systems, respectively, as well as lay the groundwork for studies of anisotropic transport phenomena such as sound propagation or lattice thermal conductivity.
\end{abstract}

\keywords{layered materials, vdW, anisotropy}

\maketitle

\section{Introduction}

Van der Waals (vdW) layered materials have served as electrodes, thermoelectrics, optoelectronics, substrates, and as precursors for 2D materials.\cite{Radisavljevic2011,Kaul2014,Novoselov,Aykol2015,Duong2017,Gorai2016b,Cheon2017,Melamed2017,Rasmussen2015,Pandey2016,Mounet2018,Singh2015,Miro2014,Novoselov2005,Radisavljevic2011} The chemistry of known vdW materials is diverse, ranging from simple unaries and binaries (e.g. graphite, SnSe, Bi$_2$Te$_3$) to complex multinaries (e.g. LiCoO$_2$, BiCuOSe).\cite{Ashton2017,Tyagi2016,Shao2016,Lee2013,Gorai2016b} Figures 1a and 1b show two example crystal structures of well-known vdW layered materials. Although individual structures have been known for decades, the classification and large-scale compilation of vdW layered structures came surprisingly late.\cite{Ashton2017,Gorai2016b,Cheon2017,Mounet2018,Naguib2014} These large scale classifications were enabled by automated identification algorithms utilizing either topology scaling\cite{Cheon2017,Mounet2018,Ashton2017} or slab cutting approaches.\cite{Gorai2016b}  Such classification enabled the high-throughput computational screening of vdW layered materials with properties tailored for thermoelectrics, hydrogen production, photocatalytic, and micro and nanoelectronic applications.\cite{Novoselov2005, Miro2014, Kaul2014, Pandey2015a, Singh2015, Gorai2016b, Ashton2017} These efforts have led to the investigation of unconventional vdW layered materials such as ZrTe$_5$,\cite{Gorai2016b} MoWSeS,\cite{Miro2014,Duong2017} and PdTe$_2$.\cite{Pandey2015a,Rasmussen2015,Kaul2014,Mounet2018}  Additionally, the classification is useful for identification of exfoliatable bulk materials which are precursors for 2D materials.\cite{Kaul2014, Miro2014, Rasmussen2015, Cheon2017, Ashton2017, Mounet2018} 

\begin{figure*}[t!]
 \centering
 \includegraphics[width=0.9\textwidth]{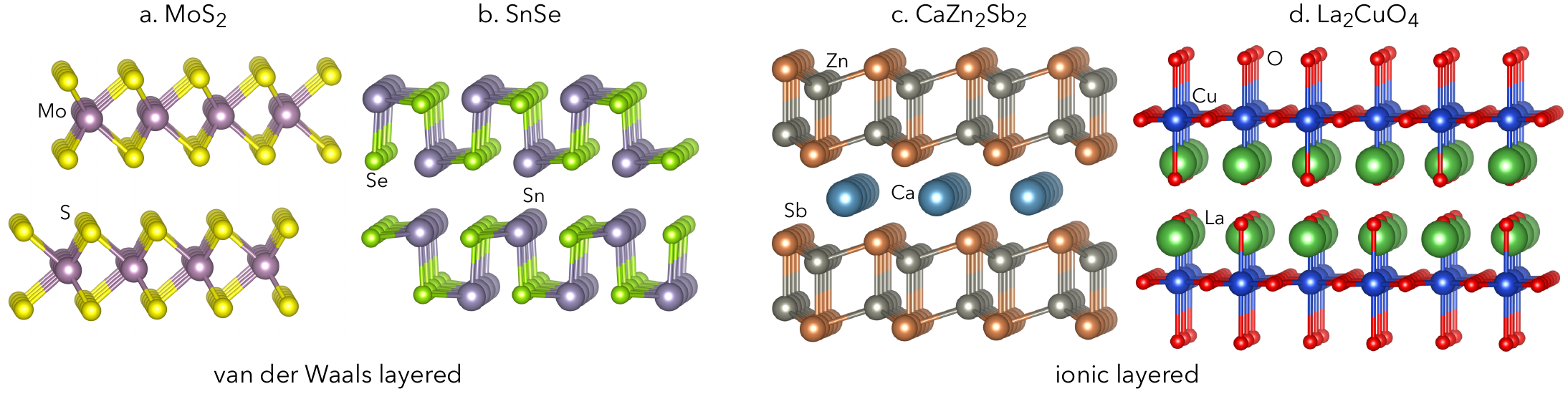}
 \caption{Crystal structures of binary vdW layered materials: (a) MoS$_2$ and (b) SnSe. Crystal structures of ternary ionic layered materials: (c) CaZn$_2$Sb$_2$, where the Mg$_2$Sb$_2$ layers are separated by Ca cations,  and (d) CuLa$_2$O$_4$, where the CuO$_4$ layers are separated by La cations. In ionic layered structures, the atoms that separate the layers (Ca, La) are referred to as ``spacers''.}
 \label{fig:LayeredExamples}
\end{figure*}

In contrast to vdW layered materials, the classification of layered materials with stronger (e.g. ionic) interlayer bonding has only recently been investigated. The singular example is the extension of the topology scaling method to screen Na-containing materials for battery applications.\cite{Zhang2018} Therein, the layers are ionically bound by the electropositive Na cation. The procedure involves selective removal of Na atoms from the structure and analysis of the pseudo-structure with the topology scaling method.\cite{Ashton2017}  This yielded 150 candidate layered structures with Na ``spacer'' atoms between the layers. In this work, we refer to the interlayer atom as the ``spacer''; hereafter, we refer to these materials as ``ionic layered'' to distinguish them from vdW layered materials.  

Beyond electrochemical applications, ionic layered materials are also known for their remarkable electronic and phonon properties.  Figures 1c and 1d highlight two example structures of ionic layered materials: CaZn$_2$Sb$_2$, which is a well-known thermoelectric material \cite{Toberer2010a,Gascoin2005,Zhang2008,Sun2017,May2012} and La$_2$CuO$_4$, a superconducting material,\cite{Labbe1987, Massidda1987, Anderson1987, Leggett1999}. Within thermoelectrics, ionic layered materials offer the opportunity to decouple electron and phonon transport owing to their inherent anisotropy.  In cuprate superconductors, such as La$_2$CuO$_4$ and similar, anisotropy also plays a critical role as the superconductivity is shown to occur within a single CuO$_2$ layer.\cite{Logvenov699} Ionic layered materials can also serve as precursors for exotic 2D materials through physical or chemical exfoliation (e.g. MAX structure to MXene structure conversion).\cite{Naguib2014,Lukatskaya2013} 

Many ternary ionic layered materials are structurally related to binary vdW layered materials through the removal of the spacers. Such pairs include TiS$_3$ and KTh$_2$Te$_6$,\cite{Papoian2000} BN and KSnSb,\cite{Yan2015,Huang2018,Ortiz2017} and Mxenes with MAX phase structure types.\cite{Lukatskaya2013, Naguib2014} The properties of these structural pairs may exhibit similar features and enable comparative studies between ionic and vdW layered materials. Such pairs may likewise enable unusual or metastable reactions through intercalation/de-intercalation of the spacer. To our knowledge, no large-scale classification of these vdW-ionic pairs has been conducted.  
  
In this paper, we utilize systematic classification of layered materials to further explore the relationship between vdW binary and ternary ionic layered materials. We begin by validating a slab cutting-based approach to identify ionic layered materials; applying this to $\sim$8,000 ternary structures from the Inorganic Crystal Structure Database (ICSD)\cite{Belsky2002} yields 1,577 ionic layered materials.  To link binary vdW structures\cite{Gorai2016b} to ternary ionic analogs, we analyze the stoichiometry, symmetry, and local coordination  of potential pairs. The analysis of the corresponding crystal structures reveal that less than half of the ternary ionic layered materials are analogs of binary vdW prototypes. To understand the connection between intra/interlayer bonding and elastic properties, we conduct a broad assessment and analysis of the elastic anisotropy of ternary ionic layered structures and previously published vdW data.\cite{Manna2018}  We find that the ionic layered materials are overall more isotropic than the vdW layered materials; however, the overall range of anisotropy spanned by both groups of layered materials is similar. We discuss several examples to both demonstrate the relationship between binary vdW layered materials and ternary ionic layered materials as well as highlight several unique manifestations of elastic anisotropy. This material dataset provides a foundation to pursue further studies of transport phenomena in layered materials.

\section{Computational Methodology}
\subsection{Identification of Ionic Layered Materials} 
%
We define a ternary ionic layered material as being formed of individual binary layers that are separated by a spacer element. If the spacer atoms are removed, the remaining pseudo-binary structure will contain relatively large spatial gaps orthogonal to the layering direction, much like in a vdW layered structure. To identify ionic layered materials from known crystal structures we utilize a previously developed algorithm used to identify vdW layered materials.\cite{Gorai2016b}

To determine if a structure is ionic layered, one needs to: ({\it i}) remove one atom type and construct the corresponding pseudo-binary structure, ({\it ii}) perform slab cutting in various directions defined by the appropriately chosen set of Miller indices $(hkl)$, ({\it iii}) for each $(hkl)$, select a termination that minimizes the undercoordination (number of broken bonds) of the surface atoms relative to their coordination in the bulk, and finally, ({\it iv}) count the number of broken bonds for every $(hkl)$. Then, if for exactly one choice of $(hkl)$ there is a termination that produces zero undercoordination of the surface atoms, the algorithm identifies the pseudo-binary structure as quasi-2D with $(hkl)$ layering direction and the removed atom type is labeled as the spacer element. If there is more than one choice of $(hkl)$, the pseudo-binary structure is of lower dimension, {\it i.e.}, for two inequivalent $(hkl)$ the structure consists of quasi-1D chains separated by spacers, or for three or more inequivalent $(hkl)$ the structure consists of isolated clusters of atoms separated by spacers (quasi-0D). Alternatively, the structure is three dimensional if the cutting routine cannot find any $(hkl)$ that fulfills the above criteria. 

\begin{figure}[!t]
\centering
\includegraphics[width=1.0\linewidth]{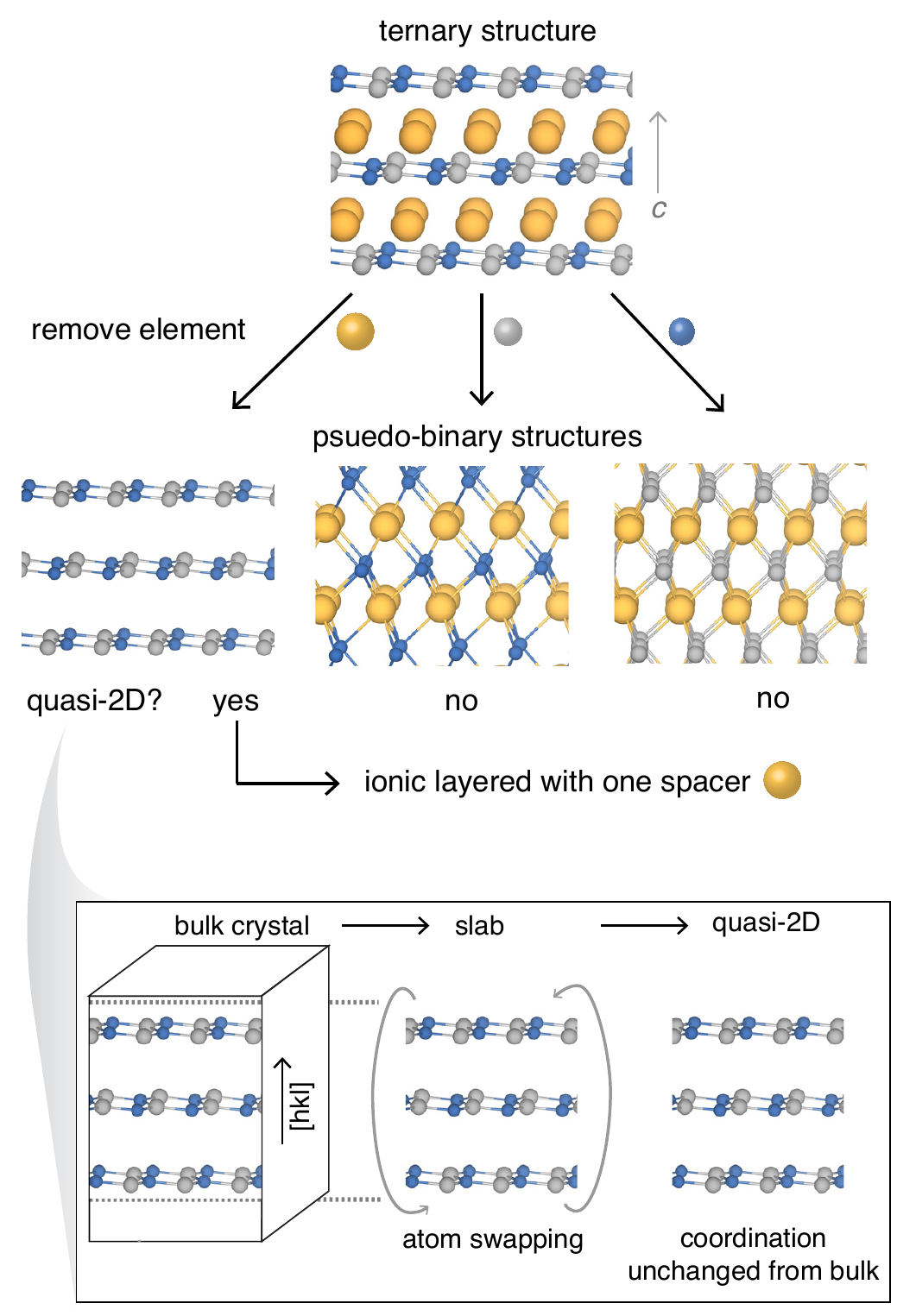}
\caption{Schematic of the algorithm for identifying ionic layered materials. For a given ternary structure, three pseudo-binary structures are generated by removing element of each type. Using a previously-developed algorithm, each of the pseudo-binary structures are classified as 3D, quasi-2D, -1D, or -0D. If one or more pseudo-binary structures are quasi-2D, the ternary compound is classified as an ionic layered material.}
\label{fig:flowchart}
\end{figure}

A schematic of the identification procedure is shown in Fig. \ref{fig:flowchart}. For a given ternary material, three pseudo-binary structures are created by removing each element in turn. For every pseudo-binary structure a slab cutting routine is applied for all symmetry inequivalent Miller indices with $-3 \leq h,k,l \leq 3$. The cutting routine is constructed specifically to find the surface termination that minimizes the undercoordination of surface atoms relative to their coordination in the bulk. This is done by interchanging positions of surface (undercoordinated) atoms from the bottom to the top of the slab and vice versa by applying the appropriate lattice vector until the undercoordination is minimized. Undercoordination is determined relative to the number of first-shell neighbors in the bulk, which is calculated by counting atoms that are separated from the central atom within a certain cutoff distance. In this way one can construct correct terminations of both flat and very corrugated surfaces. In the next step, the algorithm counts the number of $(hkl)$ sets that result in zero undercoordination and classifies the pseudo-structure accordingly. The surface cutting routine has been successfully employed to obtain realistic surface structures of various surfaces of different inorganic solids for the purpose of calculating their electronic structure and surface dipoles.\cite{Stevanovic2014a,Stevanovic2014b} 

In order to compare the ionic layered materials with the prototypical binary vdW layered materials, we constrained our search space to ternary chemistries from the ICSD.\cite{Belsky2002} In total, we considered 8,939 stoichiometric and ordered ternary materials (A$_x$B$_y$C$_z$) from the ICSD with chemistries excluding the rare earths (except La) for computational convenience.\footnote{We avoid the $f$-electron materials because DFT is known to be highly inaccurate in describing the electron correlation in $f$-electron systems and also suffers from convergence issues due to the highly localized $f$-electron states.} Out of the 8,939, our identification algorithm revealed 1,840 ternary ionic layered structures. 

In many cases, the algorithm will identify more than one element from the ternary chemistry which can act as a spacer, and consequently two possible pseudo-binary structures. In such cases, we label the most electropositive element as the primary spacer, since the majority of the known ionic layered systems occur with a cation acting as the spacer.\cite{Jiang1998,Lukatskaya2013,Aydinol1997}

\subsection{Determination of Ionic Layered-vdW Pairs} 
Because ternary ionic layered structures are sometimes known analogs of binary vdW layered materials,\cite{Papoian2000} we have created an automated algorithm that identifies this relationship. This algorithm works by comparing every identified ternary ionic material to every binary vdW layered material. 

The algorithm first removes the identified spacer to re-create the pseudo-binary structure found in the identification routine (in the case of multiple identified spacers, we use each possible pseudo-binary structure). A vdW binary and the ionic psuedo-binary structure are matched if they are equivalent on three criteria: ({\it i}) stoichiometry, ({\it ii}) space group, and ({\it iii}) average first shell coordination number. For computational efficiency, we first check the pseudo-binary structure stoichiometry against each binary vdW stoichiometry. If there is a match, we then compare the space group of the pseudo-binary structure to the vdW binary space group. Only if both the stoichiometry and the space group match do we compare the average first shell coordination number. If all three criteria are a match, we identify the binary vdW layered material as an analog of the ternary ionic layered material. 

\subsection{High-throughput Calculation of Elastic Anisotropy} 
In order to assess the effects of the different bonding mechanisms (vdW vs. ionic), we investigate the elastic properties of the binary vdW and the ternary ionic layered materials. To determine the elastic stability of each material, we use the Born stability criteria,\cite{Born1940, Karki1997, Morris2000, Mouhat2014, DeJong2015} determined from the elastic tensor. The elastic tensor ($C_{ijnm}$) is a rank-4 tensor, which relates the induced strain tensor ($\sigma_{ij}$) to an applied stress tensor ($\epsilon_{nm}$) in the following way:

\begin{align}
\sigma_{ij}=\sum_{nm}C_{ijnm}\epsilon_{nm} \label{eq:elastic}.
\end{align}

With the help of symmetry properties of the elastic tensor, the 81 component rank-4 tensor can be represented as a 6x6 matrix in the Voigt notation\cite{Hill1952}

\begin{align}
\begin{bmatrix}
    \sigma_{1}  \\
    \sigma_{2}  \\
    \sigma_{3}  \\
    \sigma_{4}  \\
    \sigma_{5}  \\
    \sigma_{6}  
\end{bmatrix}
=
\begin{bmatrix}
    C_{11} & C_{12} & C_{13} & C_{14} & C_{15} & C_{16} \\
    C_{21} & C_{22} & C_{23} & C_{24} & C_{25} & C_{26} \\
    C_{31} & C_{32} & C_{33} & C_{34} & C_{35} & C_{36} \\
    C_{41} & C_{42} & C_{43} & C_{44} & C_{45} & C_{46} \\
    C_{51} & C_{52} & C_{53} & C_{54} & C_{55} & C_{56} \\
    C_{61} & C_{62} & C_{63} & C_{64} & C_{65} & C_{66}
\end{bmatrix}
\begin{bmatrix}
    \epsilon_{1}  \\
    \epsilon_{2}  \\
    \epsilon_{3}  \\
    2\epsilon_{4}  \\
    2\epsilon_{5}  \\
    2\epsilon_{6}
\end{bmatrix}. \label{eq:voigtmatrix}
\end{align}

According to Born stability criteria,\cite{Born1940} elastically stable materials will always have positive eigenvalues of the 6$\times$6 elastic matrix, meaning that elastically stable materials always have positive elastic energy for arbitrary homogeneous deformation by an infinitesimal strain.\cite{Morris2000,Mouhat2014,DeJong2015} If there are any negative eigenvalues, the material is elastically unstable and we do not consider it in the further analysis. Elastic instabilities are associated with the point of phase transition, indicated by unphysical values of elastic parameters, such as negative Young's modulus. As such, our exclusion of such materials is warranted.\cite{Karki1997} In addition, we discard all systems that exhibit imaginary (optical mode) phonon frequencies at the $\Gamma$-point, which is an indication of dynamical instability (at the $\Gamma$-point, acoustic modes are often imaginary due to numerical precision since the values are very close to zero). Out of the 1,840 identified ternary compounds, 1,577 were found to be elastically and dynamically (at $\Gamma$) stable.

Given the layered nature of the compounds under consideration, the expectation is that many will exhibit a significant degree of anisotropy. For the purpose of assessing the anisotropy of layered systems, we evaluate the universal anisotropy index ($A_U$) for each compound.\cite{Ranganathan2008} Calculation of the universal elastic anisotropy index ($A_U$) makes use of the upper (Voigt) and lower (Reuss) bounds to the isotropic bulk and shear moduli.\cite{Ranganathan2008,Peng2015,Hill1952,Kube2016a} 
\begin{align}
A_U = \frac{B_V}{B_R} + 5\frac{G_V}{G_R} - 6 > 0.
\end{align}
This measure adopts values close to 0 for isotropic materials and values larger than 1 as those displaying pronounced elastic anisotropy. We also compute the directional Young's Modulus $E$, which represents the response of a material to axial stresses, in every direction for in order to get a sense of the directional variation of the elastic properties. The directional Young's Modulus is calculated by first finding the rank-4 compliance tensor ($S=C^{-1}$).\cite{Ting2005,Peng2015} The directional Young's Modulus along any unit vector, $\vec{u}$ can then be found by:
\begin{align}
\frac{1}{E(\vec{u})} =\sum_{ijkl} S_{ijkl}u_iu_ju_ku_l.
\end{align}
Computing the Young's Modulus over a large, evenly distributed set of directions, we find the relative variation of the Young's modulus by dividing the the angular standard deviation by the angular average, which can serve as a useful comparison metric with $A_U$.

The starting structures are obtained from the ICSD database.\cite{Belsky2002} Structural relaxation, calculation of elastic tensor and phonon frequencies (at $\Gamma$) are performed using the Vienna Ab-initio Simulation Package (VASP)\cite{Kresse1996,Kresse1996a} with projector augmented waves (PAW)\cite{Blochl1994,Kresse1999} in the generalized gradient approximation using the Perdew-Burke-Ernzerhof (PBE)\cite{Perdew1996} exchange-correlation functional. All calculations are performed with a plane wave cutoff energy of 520 eV and a fairly dense $\Gamma$-centered k-point grid of 1000 per inverse atom is used for all materials.\cite{Monkhorst1976} While the ternary ionic layered structures are relaxed with standard GGA-PBE functionals, binary vdW materials were previously relaxed using a vdW-corrected exchange correlation functional (optB86) to correctly account for the long-range vdW interactions.\cite{Gorai2016b,Klimes2011,Manna2018}  Elastic tensors are calculated with a finite difference method, in which six finite distortions of the lattice are performed and the elastic constants ($C_{ij}$) are derived from the stress-strain relationship, yielding the full elastic tensor.\cite{LePage2002,Wu2005,Manna2018} Calculations are handled within the python-based high-throughput framework, PyLada.\cite{Pylada}

\begin{figure}[!t]
\centering
\includegraphics[width=1.0\linewidth]{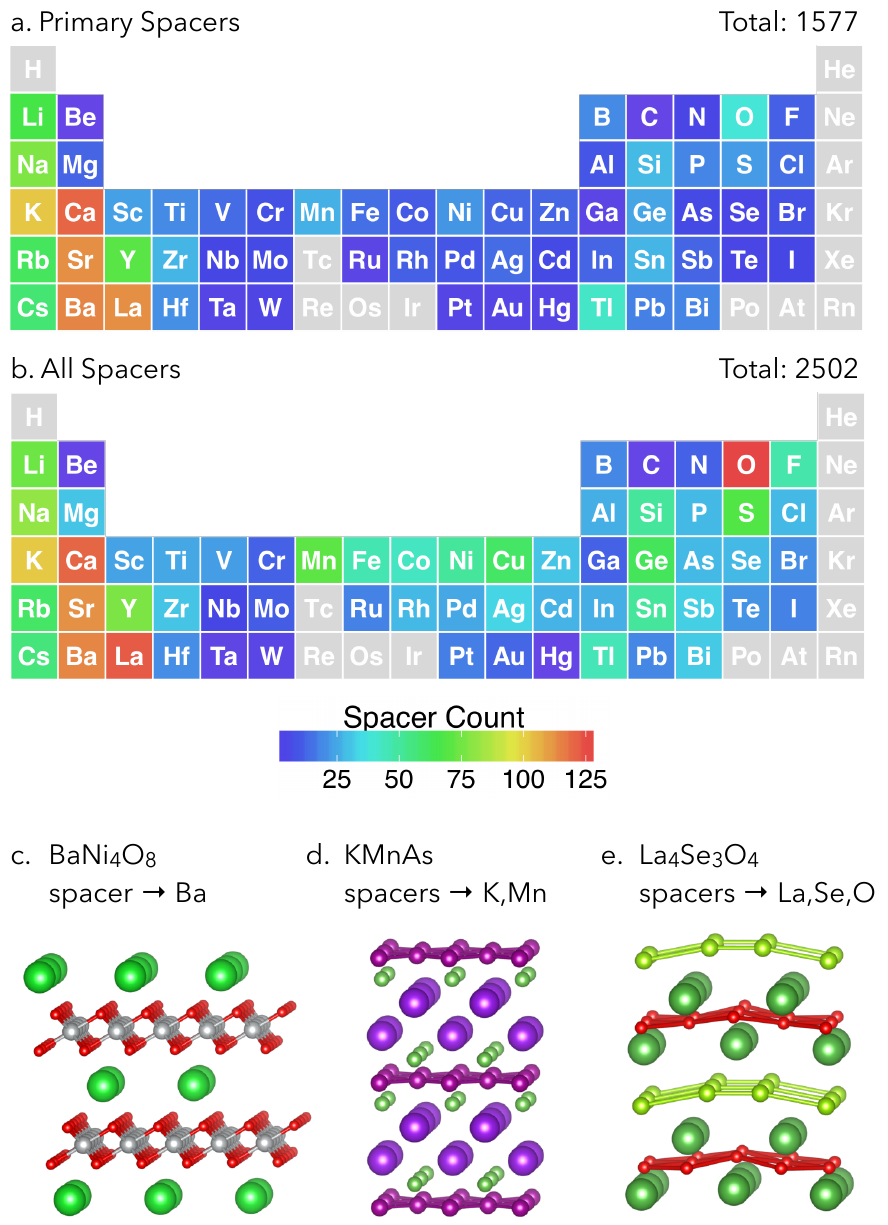}
\caption{Frequency of spacer elements. (a) Primary spacers (the most electropositive element in the case of multiple spacers). (b) All identified spacers, including all identified spacers in the case of multiple spacers. Example crystal structures with (c) one (BaNi$_4$O$_8$), (d) two (KMnAs), and (e) three  (La$_4$Se$_3$O$_4$) spacers. }
\label{fig:Ptable}
\end{figure}

\section{Results and Discussion}
Manual inspection of the search outputs indicates the ionic layered materials were successfully classified.  These ranged from classic systems (e.g. KSnSb, CaZn$_2$Sb$_2$, LiMnO2, CuLa$_2$O$_4$) to more esoteric materials (e.g. RbV$_3$O$_8$, BaClF).
Likewise, the majority of the subset of Na-containing compounds from Ref. \citenum{Zhang2018} reappear within this dataset, including NaAlSi, NaZnP, NaCoO$_2$, NaVF$_4$, Na$_2$ZrSe$_3$, and NaTiS$_2$. For systems with ambiguity in terms of their classification ({\it i.e.}, layered vs. not), the cutoff distance can lead to arbitrary classification. Materials which could be perceived as one dimensional can be misidentified as layered if the cutoff distance enables bonding between the chains. Examples include BaVS$_3$, TiMo$_6$Se$_8$, and KCuCl$_3$. Conversely, for materials where the layering is subtle, the materials may not be identified as layered (e.g. SrZn$_2$Ge$_2$, CaClF, and LaTiO$_3$). When incorrect classification was observed by visual inspection of the structure, compounds were manually reclassified. We note that all search methods constructed to use solely geometric features suffer from the ambiguity of defining the cutoff distance for bond lengths.  

\subsection{Distribution of Identified Spacer Elements}
As constructed, the slab cutting algorithm is agnostic to the chemistry of the ``spacer'' element. The chemical identity of the spacer is thus found to vary widely, as shown in Fig. \ref{fig:Ptable}. As stated previously, in the case of multiple identified spacer elements, we choose the most electropositive element as the primary spacer. In  Fig. \ref{fig:Ptable}a, we show the distribution of primary spacers. In the Fig. \ref{fig:Ptable}b, we show all identified spacers (those with two or three identified spacer elements are double and triple counted). The occurrence of multiple spacer elements is highlighted in progression through Fig. \ref{fig:Ptable}c-e.

Of the 1,577 compounds, 774 had at least two identified spacers and 151 compounds had all the three elements as spacers. We find that for the 1,577 identified compounds, 59\% of the primary spacers come from the groups I-III of the periodic table as shown in Fig. \ref{fig:Ptable}, which drops to 39\% when we include all identified spacers. The top six most frequent are Ca, Ba, Sr, La, K, and O,  each appearing over 100 times. The observation of cationic spacers as dominating is consistent with well known expectations for ionic layered materials. Many of these materials with spacers in groups I-III are Zintl or polar intermetallics (e.g. Sc$_6$PdTe$_2$, CaBe$_2$Ge$_2$).\cite{Corbett2010,Eisenmann1972} Perhaps less appreciated are the layered structures with electronegative spacers. Examples such as ClCa$_2$N and I$_3$La$_5$Si$_5$ have halogen spacers and polycationic slabs. These ternary structures tend to cap the polycationic slabs with the most electropositive element, in contrast to the binary vdW layered materials and many ternary Zintl materials. In between the highly electronegative and electropositive spacers, we were surprised to see a significant minority of transition metal and metalloid spacers. For example, ZrVSi, TlCdS$_2$, and AuCr$_3$O$_8$ are discussed below in terms of their anisotropic elasticity. 

\subsection{Space Group Prevalence}

\begin{table}[!b]
    \centering
	\begin{tabular}{lcccc} 
	\hline
	Crystal & binary & ternary & binary & ternary \\
    System & ICSD & ICSD & vdW & ionic\\
	\hline
    triclinic & 1.3 & 3.2 & 1.1 & 0.6\\
    monoclinic & 14.2 & 19.6 & 18.2 & 12.9\\
    orthorhombic & 23.2 & 27.6 & 18.2 & 24.2\\
    tetragonal & 14.8 & 16.4 & 13.0 & 31.7\\
    trigonal & 11.0 & 10.5 & 38.9 & 17.5\\
    hexagonal & 14.6 & 9.7 & 10.7 & 13.1\\
    cubic & 20.8 & 12.9 & 0.0 & 0.0\\
    \hline
	\end{tabular}
    \caption{Crystal system distribution (in percentage) of binary (3,440), ternary (8,939), binary vdW layered (3,47), and ternary ionic layered (1,577) structures reported in the ICSD (stoichiometric and ordered structures). The distribution of vdW and ionic layered structures are expressed as percentages of binary and ternary structures in the ICSD, respectively. While binary vdW compounds are predominantly found in trigonal structures, ternary ionic layered compounds are found in tetragonal structures. } 
    \label{tab:spg}
\end{table}

\begin{figure*}[!t]
\centering
\includegraphics[width=1.0
\linewidth]{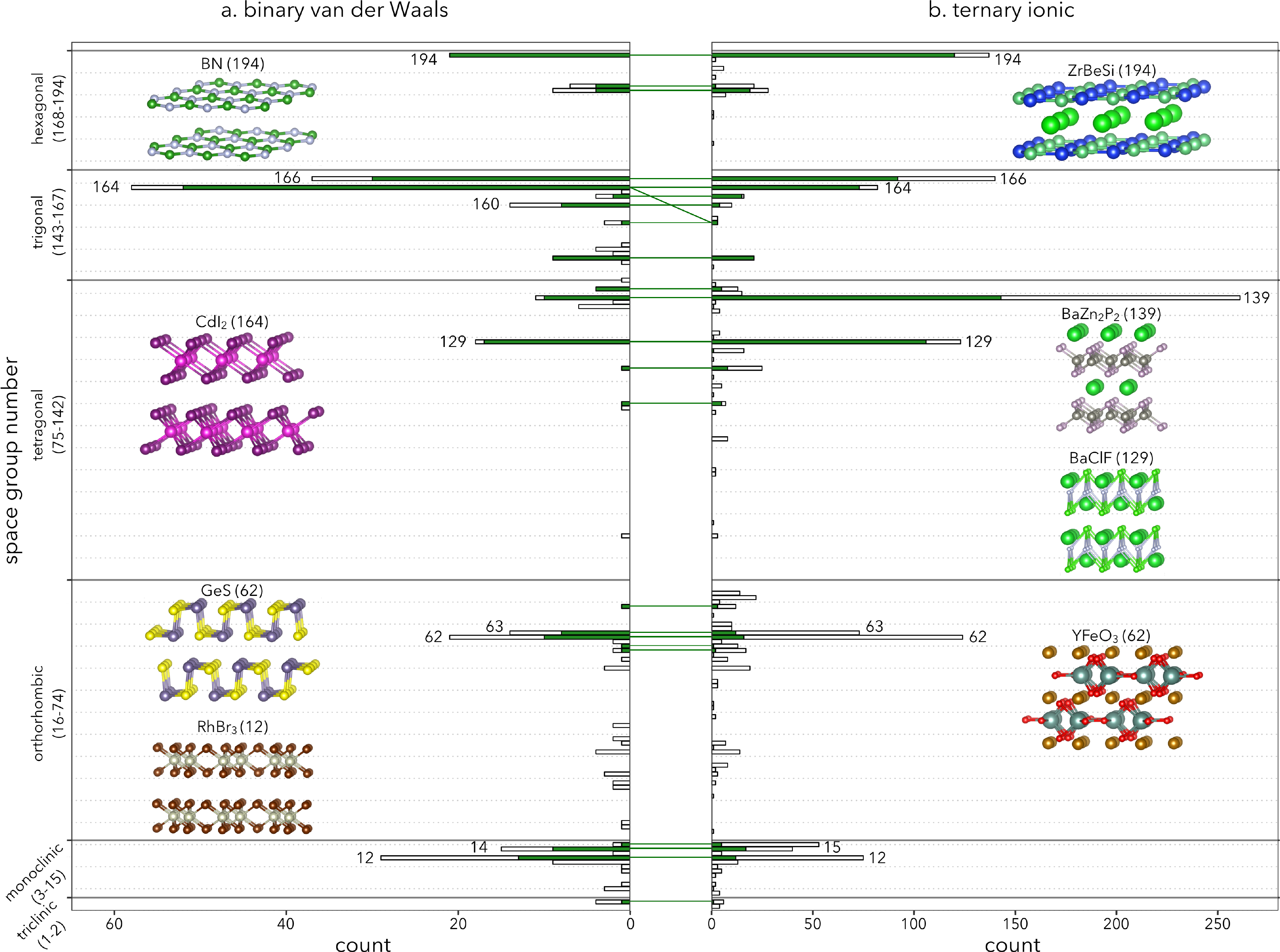}
\caption{ Histogram of space group distribution: (a) binary vdW layered, and (b) ternary ionic layered. Green bars indicate the number of materials that form vdW-ionic pairs; white bars represent those that do not form pairs. The green lines indicate the pairing between vdW and ionic layered materials. Representative crystal structures of several space groups are shown as insets.}
\label{fig:space groups}
\end{figure*}

The distribution of both sets of layered materials over the 230 space groups can be seen in Fig.~\ref{fig:space groups}. As expected, there are no cubic structures within either of these groups. The binary vdW layered compounds crystallize in 51 different space groups, but 65\% reside within the top 9 (labeled in the figure). The ternary ionic layered compounds crystallize in 84 different space groups; however, as with the binary vdW layered materials, 67\% reside with the top 9 (labeled).

The most prevalent vdW space group is the trigonal P$\overline{3}$m1 (164), with 58 out the 347 compounds. The majority of these compounds (49 of 58) are of the A$_1$B$_2$ stoichiometry, like the structure type CdI$_2$ as shown. The most populous ionic space group is the tetragonal I4/mmm (139), with  261 compounds. The A$_1$B$_2$C$_2$ stoichiometry dominates this space group, with 179 of the 261 having this structure. The corresponding structure type for these compounds is BaZn$_2$P$_2$, with Ba as the large cation spacer between the Zn$_2$P$_2$ layers. These BaZn$_2$P$_2$ compounds are different from the A$_1$B$_2$C$_2$ Zintl thermoelectrics referenced in the introduction, which are contained in the trigonal space group P$\overline{3}$m1 (164).\cite{Sun2017} 

Compounds of the PbClF structure type in space group P4/nmm (129) are an interesting example of spacer to layer bonding. BaClF, shown in Fig. \ref{fig:space groups}, has layers in which the spacer atoms are inside of the layers. The Ba atoms sit in pockets within the layer and can bond to Cl atoms directly across in the opposite layers, creating shorter interlayer bonds than were possible before the addition of the spacer element. 

A frequent space group in both sets is the hexagonal group P6$_3$/mmc (194). The vdW materials in this group consist almost entirely of binary analogs to graphite, such as BN. The ionic ternary materials in this space group are almost all intercalated BN-like structures with A$_1$B$_1$C$_1$ stoichiometry, like the structure type ZrBeSi shown in \ref{fig:space groups}. For ternary ionic layered structures in this space group which are not A$_1$B$_1$C$_1$, we find known M$_{n+1}$AX$_n$ phase materials (where M is an early transition metal, A is a group 13 or 14 element, X is C or N, and n = 1,2, or 3), such as Ti$_2$AlN and Ti$_4$AlN$_3$.\cite{Naguib2014,Barsoum2000} 

The orthorhombic space group Pnma (62) represents a relatively large class of ionic layered systems with very few vdW analogs. Most compounds in this space group exhibit the A$_1$B$_1$C$_3$ distorted perovskite structure (structure type GdFeO$_3$), many of which are known ferroelectrics\cite{Zhong1995,Fompeyrine1999}. Fig \ref{fig:space groups} shows YFeO$_3$, where Fe is the spacer atom. Here, the bonds are drawn between Y-O to highlight the layering. While it is debatable as to whether or not these perovskites should be included, we choose to leave them in because the structural distortion leads to anisotropy. For comparison, the cubic perovskites are not identified as ionic layered. Space group 62 is also frequent amongst vdW materials; however, the majority of these materials are of the GeSe structure with no analog among the ionic ternary materials.

The difference in crystal structure bias between vdW and ionic groups can be seen in Table~\ref{tab:spg}, where we report the crystal structure distribution of both sets of layered materials as compared to the binary and ternary compounds in the ICSD of which they are subsets. Binary vdW layered materials most frequently appear as trigonal structures, whereas ternary ionic layered materials are more concentrated in tetragonal structures. The key result from the structural comparison of vdW compounds to ionic compounds is that the overall distribution of structures is significantly different between the two groups, emphasizing that the ionic layered compounds are a structurally distinct class of layered compounds.

\subsection{Structural Links Between Ionic and vdW Materials}

The comparison algorithm which matches ionic compounds to vdW analogs reveals that out of 1,577 ternary ionic layered compounds, 686 have at least one binary vdW layered analog. Of the 347 binary vdW layered compounds, only 209 were found to have known ternary ionic layered derivatives. By comparing every binary vdW layered material to every ternary ionic layered material, we found 8,551 matches. These results are shown along with the distribution of space groups in Fig. \ref{fig:space groups}. In all cases, removing an atom preserves crystal symmetry. However, in two materials (AgAlS$_2$ and LiMnSe$_2$), removal of the spacer atom does change the space group from P3m1 (156) to P$\overline{3}$m1 (164). 

These relationships are intriguing synthetically.  In general, vdW materials provide inspiration for new ionic layered materials through:  (i) intercalation of known vdW compounds, yielding potentially metastable ionic materials or (ii) the vdW materials could serve as inspiration for the discovery of ionic layered materials through direct ternary synthesis that incorporates a third element to serve as the spacer and associated charge compensation within the slab. The identified vdW-ionic layered pairs provide evidence that these structural relationships are robust upon changes in bonding and can provide synthetic guidance for the discovery of new compounds. Equally intriguing are the ionic compounds that do not have a known vdW analog; such structures suggest metastable vdW compounds could be formed through de-intercalation. An example of a structure which is unique to the ionic layered ternary compounds is the La$_2$CuO$_4$ cuprate superconductor, shown in Fig.~\ref{fig:LayeredExamples}. With La removed, the CuO$_4$ layers (CuO$_2$ planes with additional O atoms above and below) that give rise to high-temperature superconductivity\cite{Anderson1987,Labbe1987,Massidda1987,Leggett1999, Logvenov699} have no analog among the vdW systems. Finally, the vdW structures that lack ionic layered pairs are particularly exciting in terms of discovering new structure types inspired by the vdW binary base structures.  

\subsection{Elastic Anisotropy}  

\begin{figure}[!t]
\centering
\includegraphics[width=0.8\linewidth]{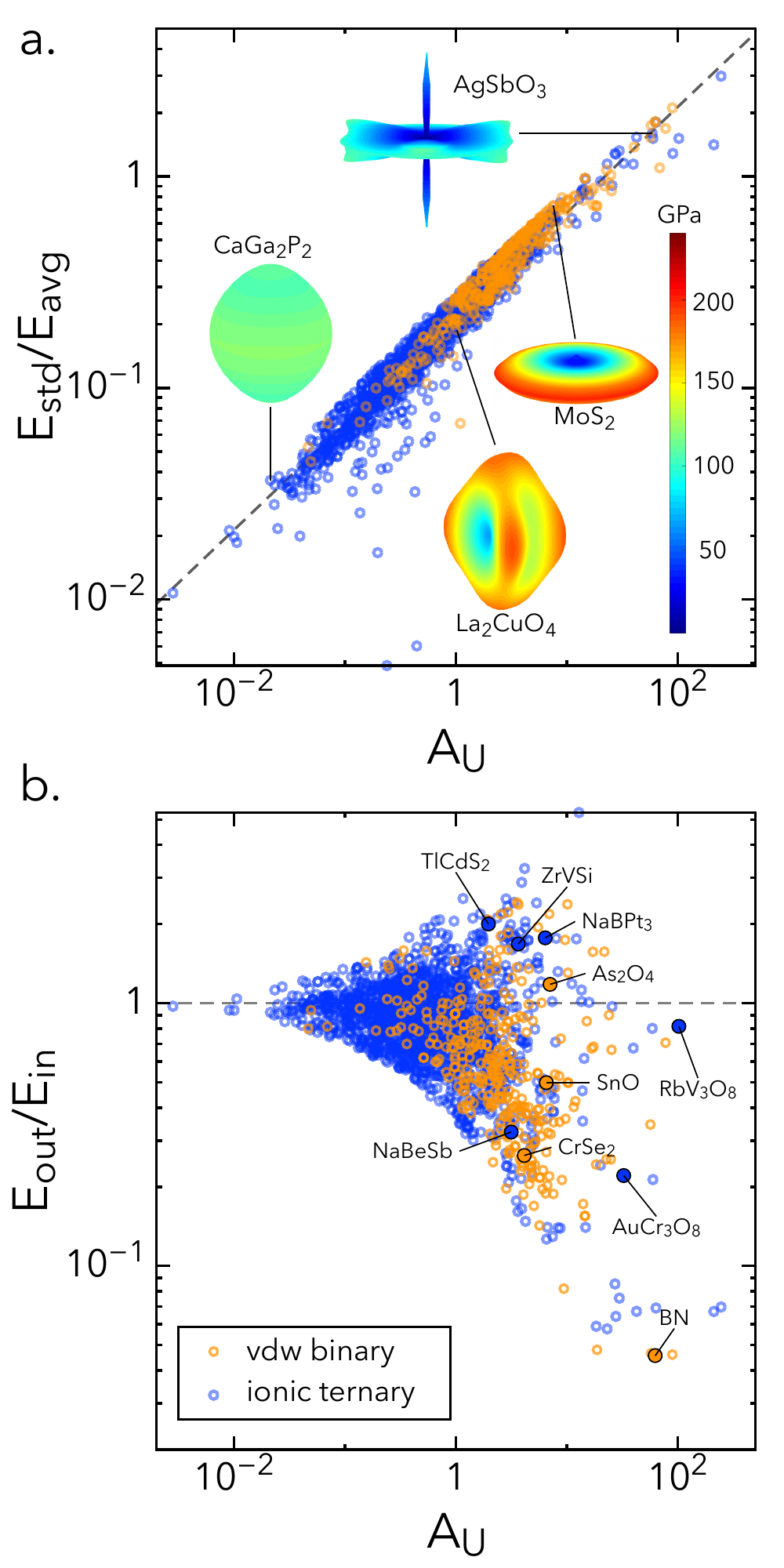}
\caption{(a) The relative standard deviation of Young's modulus as a function of the anisotropy index, $A_U$. Polar plots of Young's modulus of CaGa$_2$P$_2$, La$_2$CuO$_4$, MoS$_2$, and AgSbO$_3$ are shown as insets. Both the radius and color in the polar plots represent the magnitude (in GPa) of Young's modulus in a given direction. (b) The ratio of out-of-plane to in-plane Young's modulus ($E_{out}/E_{in}$) vs. $A_U$. Ternary ionic layered materials are shown in blue while binary vdW layered materials in orange.}
\label{fig:IonicAnisotropy}
\end{figure}

\begin{figure*}[!t]
\centering
\includegraphics[width=0.8\linewidth]{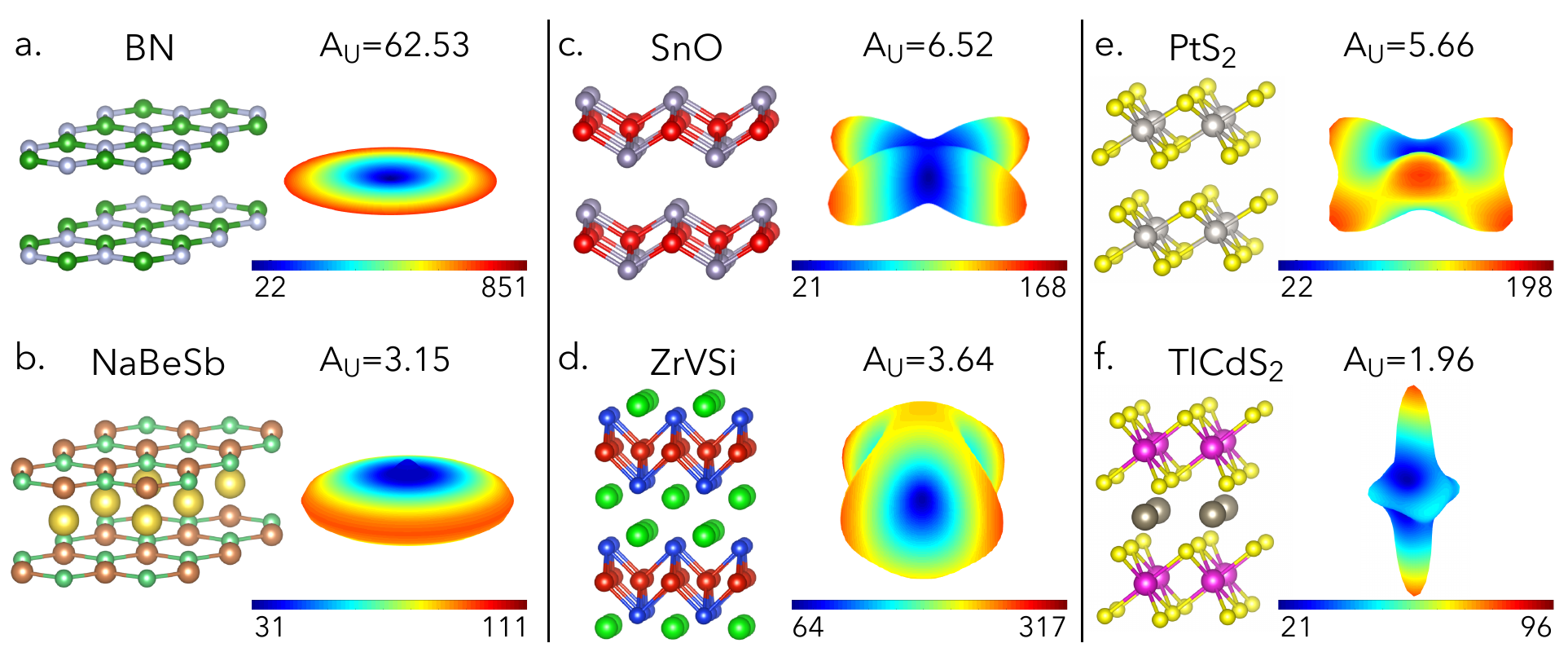}
\caption{Case examples of layered materials that form vdW-ionic pairs. For each material, the crystal structure and polar plot of Young's modulus is shown. (a-b) BN, NaBeSb, (c-d) SnO, ZrVSi, and (e-f) PtS$_2$, TlCdS$_2$. Both the radius and color in the polar plots represent the magnitude (in GPa) of Young's modulus in a given direction.}
\label{fig:strc_discussion1}
\end{figure*}

Having demonstrated the intrinsic structural differences between ionic layered ternary materials and vdW layered binary materials, we wish to understand the effect of aggregate differences in elastic properties between these two material classes. To begin, we utilize  the anisotropy index, $A_U$ as a metric for overall elastic anisotropy. In Fig. \ref{fig:IonicAnisotropy}a we assess agreement between the  anisotropy index, $A_U$, and  the relative standard deviation of the Young's modulus computed in different directions. Given the strong trend between $A_U$ and  $E_{std}/E_{avg}$, we use $A_U$ henceforth as a metric of anisotropy. By fitting the guideline in Fig.~\ref{fig:IonicAnisotropy}a to a slope of one half, we find that generally $E_{std}/E_{avg} \propto A_U^{1/2}$.

We see from Fig.~\ref{fig:IonicAnisotropy}a that $A_U$ for both ionic layered and vdW layered varies from nearly isotropic ($A_U<1$) to highly anisotropic ($A_U>10$). On average the vdW materials demonstrate a higher degree of anisotropy than the set of ionic layered materials. Only 17\% of the vdW binary layered materials have $A_U<1$, while 69\% of the ionic layered materials have $A_U<1$. Nevertheless, 484 ternary ionic layered materials show significant elastic anisotropy ($A_U>1$).  

The four example plots in Fig.~\ref{fig:IonicAnisotropy}a show the angular dependence of the Young's modulus for ionic layered CaGa$_2$P$_2$, La$_2$CuO$_4$ and AgSbO$_3$, and the well-known vdW layered hexagonal MoS$_2$. The more isotropic compound, CaGa$_2$P$_2$, with an anisotropy index of 0.02, exhibits a nearly uniform directional Young's modulus and an associated near-spherical polar plot of Young's modulus. In contrast, we find $A_U=7.55$ for MoS$_2$. The resulting Young's modulus plot of this material shows a disk-like shape with a high degree of stiffness within the layer ($xy$ plane), and very low stiffness across the layers ($z$ direction). 

In Fig. \ref{fig:IonicAnisotropy}b, we show the ratio of the Young's modulus along the layering direction ({\it i.e.}, the out-of-plane $E_{out}$) to the the average Young's modulus in the plane of the layers ($E_{in}$). As expected, materials with small A$_U$ have little difference in $E_{out}$ vs $E_{in}$. However, for systems with high A$_U$ we find little trend with $E_{out}/E_{in}$ for both vdW and ionic materials. On one extreme, NaBeSb and BN are weakly bonded between the layers. More surprising, a multitude of materials are found with stiff cross-plane bonding (e.g. TlCdS$_2$, ZrVSi). In aggregate, we find $E_{out}/E_{in}>1$ for 17.2\% and 33.2\% for vdW and ionic layered materials, respectively. The following sections consider case examples to explore the relationship between chemical structure, bonding, and anisotropy.

\subsection{Case Examples: VdW-Ionic Pairs}
We consider three of the ionic layered structures  (NaBeSb, ZrVSi, and TlCdS$_2$) for further investigation. These compounds were chosen because ($i$) they span a relatively large range of $E_{out}/E_{in}$ and ($ii$) they are analogs to the vdW layered materials BN, SnO, and PtS$_2$ respectively. The polar Young's modulus plots for these vdW and ionic layered pairs are shown in Fig. \ref{fig:strc_discussion1}. The elastic anisotropy in these and any other considered material can be understood in terms of the directions along which the axial deformations directly affect the length of chemical bonds (stiff), angles between chemical bonds (less stiff) or imply rigid rotations of bonding patterns (least stiff).

Hexagonal boron nitride (h-BN) is known for its relatively strong in-plane B-N chemical bonds leading to a pronounced anisotropy in the Young's modulus. The corresponding polar plot is an extremely thin disk, reflecting the large difference between the intralayer and the interlayer bonding ($E_{out}/E_{in}$ of 0.04) as well as in-plane isotropy of the honeycomb lattice. The ternary ionic layered analog, NaBeSb (KZnAs structure type, shown below BN), contains  hexagonal sheets of BeSb separated by Na ion spacers. As we see with the Young's modulus plot, there is little difference in the qualitative nature of the angular dependence. However, the ionic bonding between layers leads to significant increase in $E_{out}/E_{in}$ to 0.3. Considering all compounds of the KZnAs structure type, we find a $E_{out}/E_{in}$ range of 0.3 to 2.1 (NaBeSb to LaPdSb). This illustrates an important feature of the ionic layered materials: that they can allow significant chemical tuning of the in-plane vs out-of-plane elastic anisotropy, which is more difficult to achieve in vdW systems. 

\begin{figure}[!t]
\centering
\includegraphics[width=0.8\linewidth]{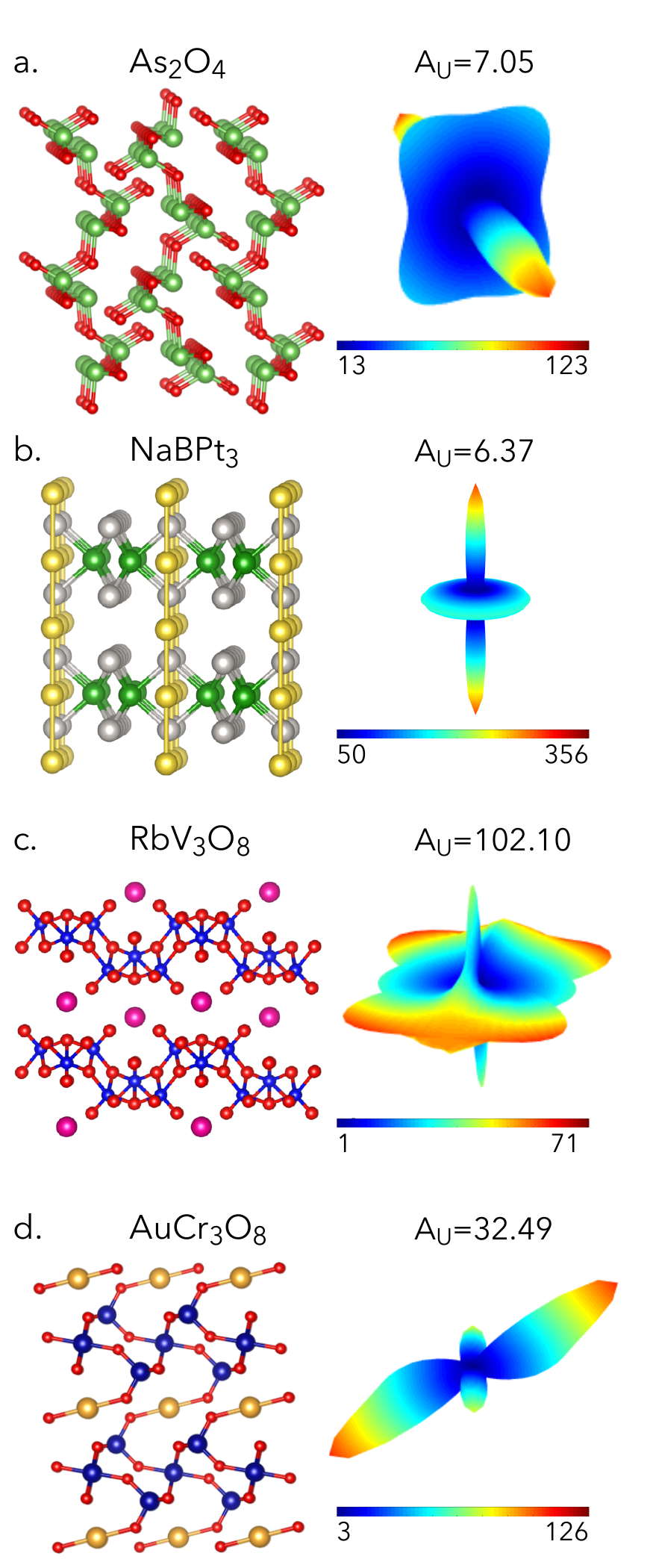}
\caption{Case examples of layered materials that do not have the corresponding vdW or ionic pairs. For each material, the crystal structure and polar plot of Young's modulus is shown. Examples of binary vdW layered (a) As$_2$O$_4$, and ternary ionic layered (b-d) NaBPt$_3$, RbV$_3$O$_8$, and AuCr$_3$O$_8$, are shown.  Both the radius and color in the polar plots represent the magnitude (in GPa) of Young's modulus in a given direction.}
\label{fig:strc_discussion2}
\end{figure}

For the next case example, we investigate tetragonal ZrVSi (ZrSSi structure type) and its vdW pair SnO to consider the impact of in-plane bonding on elastic properties. Here, the slab is built from planar, square nets of O or V, decorated above and below by Sn and Si, respectively. For SnO, compression in-plane along the O-O nearest neighbor directions yields Coulombic repulsion between the oxygen anions as the bond length shrinks, yielding directions of maximum stiffness. In contrast, compression along diagonals predominately changes bond angles between Sn-O rather than bond lengths and results in a far softer elastic response. Similar behavior in-plane is seen for ZrVSi; however, the stronger out-of-plane bonding between the layers in ZrVSi causes the Young's modulus to approach the in-plane values, unlike in the analogous vdW system.  

In the TlCdS$_2$-PtS$_2$ pairing, both materials have in-plane elastic behavior which reflect the bonding angles between the transition metal, Cd or Pt, and sulfur. The directions of maximum stiffness for PtS$_2$ lie along these six directions, since compression along any of these directions changes the Pt-S bond length. This same in-plane pattern can be seen in TlCd$_2$; however, Tl is also 6-fold bonded to sulfur, leading to strong interlayer bonding. In this case, the out-of-plane Young's modulus is significantly stronger than the in-plane values, yielding an unusual layered material which is much stiffer across the layers than in the plane of the layers. 

\subsection{Case Examples: Structures Without Pairs}

To highlight the diversity of anisotropic elasticity in layered materials, we consider several additional case examples. Fig.~\ref{fig:strc_discussion2} shows one vdW material (As$_2$O$_4$) and three ionic materials (NaBPt$_3$, RbV$_3$O$_8$ and AuCr$_3$O$_8$) which are unique layered materials without analog in the other set. The main features of the directional behavior of Young's modulus in these materials indicate complex and counterintuitive elastic anisotropy. 

We start by choosing a highly corrugated vdW layered binary, As$_2$O$_4$. This material exhibits a ``needle''-like Young's modulus polar plot, with one direction exhibiting a maximum value. However, the direction of high stiffness is perpendicular to the layering direction, along the ridges of the corrugations. This Young's modulus plot has very low values both across the layers, as expected, but also across the corrugations. 

NaBPt$_3$ also exhibits a ``needle''-like polar Young's modulus plot. In this case, the direction of maximum stiffness is perpendicular to the plane of the layers, which is due to the spacer-spacer (Na-Na) bonds that exist both between the layers and within the layers. Namely, in this structure the chains of Na atoms pass through the hexagonal holes in the BPt$_3$ layers, creating quasi-1D chains of Na passing through the BPt$_3$ layers. 

In the RbV$_3$O$_8$ and AuCr$_3$O$_8$ the more complex layer structures in combination with significant spacer-layer interactions lead to very complex and unique directional behavior of their Young's moduli. In the Young's modulus plots for both RbV$_3$O$_8$ and AuCr$_3$O$_8$, we see an out-of-plane crest directed along an angle corresponding to the bonding angle between the spacer element and oxygen. The in-plane elastic behavior is highly anisotropic for both materials. In RbV$_3$O$_8$, the layers are corrugated sheets, leading to a stiff direction along the ridge of the corrugations and a soft direction across the corrugations. In AuCr$_3$O$_8$, the Cr$_3$O$_8$, the layers are formed from very loosely bound octrahedra and tetrahedra, leading to extremely soft in-plane elasticity.

\section{Conclusions}
In this work, we have significantly expanded the set of identified layered materials through the addition of the ternary ionic layered compounds. Additionally, we have structurally matched ternary ionic layered materials to their binary vdW pairs. Where links exist, chemical tuning of the spacer element can be lead to the desired elastic anisotropy. The absence of a link presents opportunities for exploration of new materials obtained by insertion or removal of spacer elements in vdW and ionic layered materials, respectively. Our calculations reveal a diverse range of elastic anisotropy in both vdW and ionic layered materials. We show that the majority of ionic layered materials are elastically more isotropic than vdW layered materials. Nevertheless, a large population of ternary ionic layered materials exhibit highly anisotropic elastic properties, which make them interesting candidates for thermoelectric and superconducting applications. Further analysis of the elastic properties, in conjunction with the structural and chemical features, using techniques such as machine learning will likely throw light on the fundamental driving factors for anisotropy in these classes of layered materials. This work lays the foundation for further exploration of anisotropy in application-relevant properties such as thermal conductivity and carrier mobilities.

\textbf{Acknowledgement} We acknowledge support from the National Science Foundation DMR program, grant no. 1334713, and the Research Corporation for Scientific Advancement \textit{via} the Cottrell Scholar Award. The research was performed using computational resources sponsored by the Department of Energy's Office of Energy Efficiency and Renewable Energy and located at the NREL.

\bibliography{Ionic}

\begin{thebibliography}{68}%
\makeatletter
\providecommand \@ifxundefined [1]{%
 \@ifx{#1\undefined}
}%
\providecommand \@ifnum [1]{%
 \ifnum #1\expandafter \@firstoftwo
 \else \expandafter \@secondoftwo
 \fi
}%
\providecommand \@ifx [1]{%
 \ifx #1\expandafter \@firstoftwo
 \else \expandafter \@secondoftwo
 \fi
}%
\providecommand \natexlab [1]{#1}%
\providecommand \enquote  [1]{``#1''}%
\providecommand \bibnamefont  [1]{#1}%
\providecommand \bibfnamefont [1]{#1}%
\providecommand \citenamefont [1]{#1}%
\providecommand \href@noop [0]{\@secondoftwo}%
\providecommand \href [0]{\begingroup \@sanitize@url \@href}%
\providecommand \@href[1]{\@@startlink{#1}\@@href}%
\providecommand \@@href[1]{\endgroup#1\@@endlink}%
\providecommand \@sanitize@url [0]{\catcode `\\12\catcode `\$12\catcode
  `\&12\catcode `\#12\catcode `\^12\catcode `\_12\catcode `\%12\relax}%
\providecommand \@@startlink[1]{}%
\providecommand \@@endlink[0]{}%
\providecommand \url  [0]{\begingroup\@sanitize@url \@url }%
\providecommand \@url [1]{\endgroup\@href {#1}{\urlprefix }}%
\providecommand \urlprefix  [0]{URL }%
\providecommand \Eprint [0]{\href }%
\providecommand \doibase [0]{http://dx.doi.org/}%
\providecommand \selectlanguage [0]{\@gobble}%
\providecommand \bibinfo  [0]{\@secondoftwo}%
\providecommand \bibfield  [0]{\@secondoftwo}%
\providecommand \translation [1]{[#1]}%
\providecommand \BibitemOpen [0]{}%
\providecommand \bibitemStop [0]{}%
\providecommand \bibitemNoStop [0]{.\EOS\space}%
\providecommand \EOS [0]{\spacefactor3000\relax}%
\providecommand \BibitemShut  [1]{\csname bibitem#1\endcsname}%
\let\auto@bib@innerbib\@empty
\bibitem [{\citenamefont {Radisavljevic}\ \emph {et~al.}(2011)\citenamefont
  {Radisavljevic}, \citenamefont {Radenovic}, \citenamefont {Brivio},
  \citenamefont {Giacometti},\ and\ \citenamefont {Kis}}]{Radisavljevic2011}%
  \BibitemOpen
  \bibfield  {author} {\bibinfo {author} {\bibfnamefont {B.}~\bibnamefont
  {Radisavljevic}}, \bibinfo {author} {\bibfnamefont {A.}~\bibnamefont
  {Radenovic}}, \bibinfo {author} {\bibfnamefont {J.}~\bibnamefont {Brivio}},
  \bibinfo {author} {\bibfnamefont {V.}~\bibnamefont {Giacometti}}, \ and\
  \bibinfo {author} {\bibfnamefont {A.}~\bibnamefont {Kis}},\ }\href {\doibase
  10.1038/nnano.2010.279} {\bibfield  {journal} {\bibinfo  {journal} {Nat.
  Nanotechnology}\ }\textbf {\bibinfo {volume} {6}},\ \bibinfo {pages} {147}
  (\bibinfo {year} {2011})}\BibitemShut {NoStop}%
\bibitem [{\citenamefont {Kaul}(2014)}]{Kaul2014}%
  \BibitemOpen
  \bibfield  {author} {\bibinfo {author} {\bibfnamefont {A.~B.}\ \bibnamefont
  {Kaul}},\ }\href {\doibase 10.1557/jmr.2014.6} {\bibfield  {journal}
  {\bibinfo  {journal} {J. Mater. Research}\ }\textbf {\bibinfo {volume}
  {29}},\ \bibinfo {pages} {348} (\bibinfo {year} {2014})}\BibitemShut
  {NoStop}%
\bibitem [{\citenamefont {Novoselov}\ \emph {et~al.}(2016)\citenamefont
  {Novoselov}, \citenamefont {Mishchenko}, \citenamefont {Carvalho},\ and\
  \citenamefont {{Castro Neto}}}]{Novoselov}%
  \BibitemOpen
  \bibfield  {author} {\bibinfo {author} {\bibfnamefont {K.~S.}\ \bibnamefont
  {Novoselov}}, \bibinfo {author} {\bibfnamefont {A.}~\bibnamefont
  {Mishchenko}}, \bibinfo {author} {\bibfnamefont {A.}~\bibnamefont
  {Carvalho}}, \ and\ \bibinfo {author} {\bibfnamefont {A.~H.}\ \bibnamefont
  {{Castro Neto}}},\ }\href {\doibase 10.1126/science.aac9439} {\bibfield
  {journal} {\bibinfo  {journal} {Science}\ }\textbf {\bibinfo {volume}
  {353}},\ \bibinfo {pages} {aac9439} (\bibinfo {year} {2016})}\BibitemShut
  {NoStop}%
\bibitem [{\citenamefont {Aykol}\ \emph {et~al.}(2015)\citenamefont {Aykol},
  \citenamefont {Kim},\ and\ \citenamefont {Wolverton}}]{Aykol2015}%
  \BibitemOpen
  \bibfield  {author} {\bibinfo {author} {\bibfnamefont {M.}~\bibnamefont
  {Aykol}}, \bibinfo {author} {\bibfnamefont {S.}~\bibnamefont {Kim}}, \ and\
  \bibinfo {author} {\bibfnamefont {C.}~\bibnamefont {Wolverton}},\ }\href
  {\doibase 10.1021/acs.jpcc.5b06240} {\bibfield  {journal} {\bibinfo
  {journal} {J. Phys. Chem. C}\ }\textbf {\bibinfo {volume} {119}},\ \bibinfo
  {pages} {19053} (\bibinfo {year} {2015})}\BibitemShut {NoStop}%
\bibitem [{\citenamefont {Duong}\ \emph {et~al.}(2017)\citenamefont {Duong},
  \citenamefont {Yun},\ and\ \citenamefont {Lee}}]{Duong2017}%
  \BibitemOpen
  \bibfield  {author} {\bibinfo {author} {\bibfnamefont {D.~L.}\ \bibnamefont
  {Duong}}, \bibinfo {author} {\bibfnamefont {S.~J.}\ \bibnamefont {Yun}}, \
  and\ \bibinfo {author} {\bibfnamefont {Y.~H.}\ \bibnamefont {Lee}},\ }\href
  {\doibase 10.1021/acsnano.7b07436} {\bibfield  {journal} {\bibinfo  {journal}
  {ACS Nano}\ }\textbf {\bibinfo {volume} {11}},\ \bibinfo {pages} {11803}
  (\bibinfo {year} {2017})}\BibitemShut {NoStop}%
\bibitem [{\citenamefont {Gorai}\ \emph {et~al.}(2016)\citenamefont {Gorai},
  \citenamefont {Toberer},\ and\ \citenamefont
  {Stevanovi{\'{c}}}}]{Gorai2016b}%
  \BibitemOpen
  \bibfield  {author} {\bibinfo {author} {\bibfnamefont {P.}~\bibnamefont
  {Gorai}}, \bibinfo {author} {\bibfnamefont {E.~S.}\ \bibnamefont {Toberer}},
  \ and\ \bibinfo {author} {\bibfnamefont {V.}~\bibnamefont
  {Stevanovi{\'{c}}}},\ }\href {\doibase 10.1039/C6TA04121C} {\bibfield
  {journal} {\bibinfo  {journal} {J. Mater. Chem. A}\ }\textbf {\bibinfo
  {volume} {4}},\ \bibinfo {pages} {11110} (\bibinfo {year}
  {2016})}\BibitemShut {NoStop}%
\bibitem [{\citenamefont {Cheon}\ \emph {et~al.}(2017)\citenamefont {Cheon},
  \citenamefont {Duerloo}, \citenamefont {Sendek}, \citenamefont {Porter},
  \citenamefont {Chen},\ and\ \citenamefont {Reed}}]{Cheon2017}%
  \BibitemOpen
  \bibfield  {author} {\bibinfo {author} {\bibfnamefont {G.}~\bibnamefont
  {Cheon}}, \bibinfo {author} {\bibfnamefont {K.-A.~N.}\ \bibnamefont
  {Duerloo}}, \bibinfo {author} {\bibfnamefont {A.~D.}\ \bibnamefont {Sendek}},
  \bibinfo {author} {\bibfnamefont {C.}~\bibnamefont {Porter}}, \bibinfo
  {author} {\bibfnamefont {Y.}~\bibnamefont {Chen}}, \ and\ \bibinfo {author}
  {\bibfnamefont {E.~J.}\ \bibnamefont {Reed}},\ }\href {\doibase
  10.1021/acs.nanolett.6b05229} {\bibfield  {journal} {\bibinfo  {journal}
  {Nano Letters}\ }\textbf {\bibinfo {volume} {17}},\ \bibinfo {pages} {1915}
  (\bibinfo {year} {2017})}\BibitemShut {NoStop}%
\bibitem [{\citenamefont {Melamed}\ \emph {et~al.}(2017)\citenamefont
  {Melamed}, \citenamefont {Ortiz}, \citenamefont {Gorai}, \citenamefont
  {Martinez}, \citenamefont {McMahon}, \citenamefont {Miller}, \citenamefont
  {Stevanovi{\'{c}}}, \citenamefont {Tamboli}, \citenamefont {Norman},\ and\
  \citenamefont {Toberer}}]{Melamed2017}%
  \BibitemOpen
  \bibfield  {author} {\bibinfo {author} {\bibfnamefont {C.~L.}\ \bibnamefont
  {Melamed}}, \bibinfo {author} {\bibfnamefont {B.~R.}\ \bibnamefont {Ortiz}},
  \bibinfo {author} {\bibfnamefont {P.}~\bibnamefont {Gorai}}, \bibinfo
  {author} {\bibfnamefont {A.~D.}\ \bibnamefont {Martinez}}, \bibinfo {author}
  {\bibfnamefont {W.~E.}\ \bibnamefont {McMahon}}, \bibinfo {author}
  {\bibfnamefont {E.~M.}\ \bibnamefont {Miller}}, \bibinfo {author}
  {\bibfnamefont {V.}~\bibnamefont {Stevanovi{\'{c}}}}, \bibinfo {author}
  {\bibfnamefont {A.~C.}\ \bibnamefont {Tamboli}}, \bibinfo {author}
  {\bibfnamefont {A.~G.}\ \bibnamefont {Norman}}, \ and\ \bibinfo {author}
  {\bibfnamefont {E.~S.}\ \bibnamefont {Toberer}},\ }\href {\doibase
  10.1021/acs.chemmater.7b03198} {\bibfield  {journal} {\bibinfo  {journal}
  {Chem. Mater.}\ }\textbf {\bibinfo {volume} {29}},\ \bibinfo {pages} {8472}
  (\bibinfo {year} {2017})}\BibitemShut {NoStop}%
\bibitem [{\citenamefont {Rasmussen}\ and\ \citenamefont
  {Thygesen}(2015)}]{Rasmussen2015}%
  \BibitemOpen
  \bibfield  {author} {\bibinfo {author} {\bibfnamefont {F.~A.}\ \bibnamefont
  {Rasmussen}}\ and\ \bibinfo {author} {\bibfnamefont {K.~S.}\ \bibnamefont
  {Thygesen}},\ }\href {\doibase 10.1021/acs.jpcc.5b02950} {\bibfield
  {journal} {\bibinfo  {journal} {J. Phys. Chem. C}\ }\textbf {\bibinfo
  {volume} {119}},\ \bibinfo {pages} {13169} (\bibinfo {year}
  {2015})}\BibitemShut {NoStop}%
\bibitem [{\citenamefont {Pandey}\ \emph {et~al.}(2016)\citenamefont {Pandey},
  \citenamefont {Rasmussen}, \citenamefont {Kuhar}, \citenamefont {Olsen},
  \citenamefont {Jacobsen},\ and\ \citenamefont {Thygesen}}]{Pandey2016}%
  \BibitemOpen
  \bibfield  {author} {\bibinfo {author} {\bibfnamefont {M.}~\bibnamefont
  {Pandey}}, \bibinfo {author} {\bibfnamefont {F.~A.}\ \bibnamefont
  {Rasmussen}}, \bibinfo {author} {\bibfnamefont {K.}~\bibnamefont {Kuhar}},
  \bibinfo {author} {\bibfnamefont {T.}~\bibnamefont {Olsen}}, \bibinfo
  {author} {\bibfnamefont {K.~W.}\ \bibnamefont {Jacobsen}}, \ and\ \bibinfo
  {author} {\bibfnamefont {K.~S.}\ \bibnamefont {Thygesen}},\ }\href {\doibase
  10.1021/acs.nanolett.5b04513} {\bibfield  {journal} {\bibinfo  {journal}
  {Nano Letters}\ }\textbf {\bibinfo {volume} {16}},\ \bibinfo {pages} {2234}
  (\bibinfo {year} {2016})}\BibitemShut {NoStop}%
\bibitem [{\citenamefont {Mounet}\ \emph {et~al.}(2018)\citenamefont {Mounet},
  \citenamefont {Gibertini}, \citenamefont {Schwaller}, \citenamefont {Campi},
  \citenamefont {Merkys}, \citenamefont {Marrazzo}, \citenamefont {Sohier},
  \citenamefont {Castelli}, \citenamefont {Cepellotti}, \citenamefont {Pizzi},\
  and\ \citenamefont {Marzari}}]{Mounet2018}%
  \BibitemOpen
  \bibfield  {author} {\bibinfo {author} {\bibfnamefont {N.}~\bibnamefont
  {Mounet}}, \bibinfo {author} {\bibfnamefont {M.}~\bibnamefont {Gibertini}},
  \bibinfo {author} {\bibfnamefont {P.}~\bibnamefont {Schwaller}}, \bibinfo
  {author} {\bibfnamefont {D.}~\bibnamefont {Campi}}, \bibinfo {author}
  {\bibfnamefont {A.}~\bibnamefont {Merkys}}, \bibinfo {author} {\bibfnamefont
  {A.}~\bibnamefont {Marrazzo}}, \bibinfo {author} {\bibfnamefont
  {T.}~\bibnamefont {Sohier}}, \bibinfo {author} {\bibfnamefont {I.~E.}\
  \bibnamefont {Castelli}}, \bibinfo {author} {\bibfnamefont {A.}~\bibnamefont
  {Cepellotti}}, \bibinfo {author} {\bibfnamefont {G.}~\bibnamefont {Pizzi}}, \
  and\ \bibinfo {author} {\bibfnamefont {N.}~\bibnamefont {Marzari}},\ }\href
  {\doibase 10.1038/s41565-017-0035-5} {\bibfield  {journal} {\bibinfo
  {journal} {Nat. Nanotechnology}\ }\textbf {\bibinfo {volume} {13}},\ \bibinfo
  {pages} {246} (\bibinfo {year} {2018})}\BibitemShut {NoStop}%
\bibitem [{\citenamefont {Singh}\ \emph {et~al.}(2015)\citenamefont {Singh},
  \citenamefont {Mathew}, \citenamefont {Zhuang},\ and\ \citenamefont
  {Hennig}}]{Singh2015}%
  \BibitemOpen
  \bibfield  {author} {\bibinfo {author} {\bibfnamefont {A.~K.}\ \bibnamefont
  {Singh}}, \bibinfo {author} {\bibfnamefont {K.}~\bibnamefont {Mathew}},
  \bibinfo {author} {\bibfnamefont {H.~L.}\ \bibnamefont {Zhuang}}, \ and\
  \bibinfo {author} {\bibfnamefont {R.~G.}\ \bibnamefont {Hennig}},\ }\href
  {\doibase 10.1021/jz502646d} {\bibfield  {journal} {\bibinfo  {journal} {J.
  Phys. Chem. Lett.}\ }\textbf {\bibinfo {volume} {6}},\ \bibinfo {pages}
  {1087} (\bibinfo {year} {2015})}\BibitemShut {NoStop}%
\bibitem [{\citenamefont {Mir{\'{o}}}\ \emph {et~al.}(2014)\citenamefont
  {Mir{\'{o}}}, \citenamefont {Audiffred},\ and\ \citenamefont
  {Heine}}]{Miro2014}%
  \BibitemOpen
  \bibfield  {author} {\bibinfo {author} {\bibfnamefont {P.}~\bibnamefont
  {Mir{\'{o}}}}, \bibinfo {author} {\bibfnamefont {M.}~\bibnamefont
  {Audiffred}}, \ and\ \bibinfo {author} {\bibfnamefont {T.}~\bibnamefont
  {Heine}},\ }\href {\doibase 10.1039/C4CS00102H} {\bibfield  {journal}
  {\bibinfo  {journal} {Chem. Soc. Rev.}\ }\textbf {\bibinfo {volume} {43}},\
  \bibinfo {pages} {6537} (\bibinfo {year} {2014})}\BibitemShut {NoStop}%
\bibitem [{\citenamefont {Novoselov}\ \emph {et~al.}(2005)\citenamefont
  {Novoselov}, \citenamefont {Jiang}, \citenamefont {Schedin}, \citenamefont
  {Booth}, \citenamefont {Khotkevich}, \citenamefont {Morozov},\ and\
  \citenamefont {Geim}}]{Novoselov2005}%
  \BibitemOpen
  \bibfield  {author} {\bibinfo {author} {\bibfnamefont {K.~S.}\ \bibnamefont
  {Novoselov}}, \bibinfo {author} {\bibfnamefont {D.}~\bibnamefont {Jiang}},
  \bibinfo {author} {\bibfnamefont {F.}~\bibnamefont {Schedin}}, \bibinfo
  {author} {\bibfnamefont {T.~J.}\ \bibnamefont {Booth}}, \bibinfo {author}
  {\bibfnamefont {V.~V.}\ \bibnamefont {Khotkevich}}, \bibinfo {author}
  {\bibfnamefont {S.~V.}\ \bibnamefont {Morozov}}, \ and\ \bibinfo {author}
  {\bibfnamefont {A.~K.}\ \bibnamefont {Geim}},\ }\href {\doibase
  10.1073/pnas.0502848102} {\bibfield  {journal} {\bibinfo  {journal} {Proc.
  Nat. Acad. Sciences}\ }\textbf {\bibinfo {volume} {102}},\ \bibinfo {pages}
  {10451} (\bibinfo {year} {2005})}\BibitemShut {NoStop}%
\bibitem [{\citenamefont {Ashton}\ \emph {et~al.}(2017)\citenamefont {Ashton},
  \citenamefont {Paul}, \citenamefont {Sinnott},\ and\ \citenamefont
  {Hennig}}]{Ashton2017}%
  \BibitemOpen
  \bibfield  {author} {\bibinfo {author} {\bibfnamefont {M.}~\bibnamefont
  {Ashton}}, \bibinfo {author} {\bibfnamefont {J.}~\bibnamefont {Paul}},
  \bibinfo {author} {\bibfnamefont {S.~B.}\ \bibnamefont {Sinnott}}, \ and\
  \bibinfo {author} {\bibfnamefont {R.~G.}\ \bibnamefont {Hennig}},\ }\href
  {\doibase 10.1103/PhysRevLett.118.106101} {\bibfield  {journal} {\bibinfo
  {journal} {Phys. Rev. Lett.}\ }\textbf {\bibinfo {volume} {118}},\ \bibinfo
  {pages} {106101} (\bibinfo {year} {2017})}\BibitemShut {NoStop}%
\bibitem [{\citenamefont {Tyagi}\ \emph {et~al.}(2016)\citenamefont {Tyagi},
  \citenamefont {Gahtori}, \citenamefont {Bathula}, \citenamefont {Singh},
  \citenamefont {Bishnoi}, \citenamefont {Auluck}, \citenamefont {Srivastava},\
  and\ \citenamefont {Dhar}}]{Tyagi2016}%
  \BibitemOpen
  \bibfield  {author} {\bibinfo {author} {\bibfnamefont {K.}~\bibnamefont
  {Tyagi}}, \bibinfo {author} {\bibfnamefont {B.}~\bibnamefont {Gahtori}},
  \bibinfo {author} {\bibfnamefont {S.}~\bibnamefont {Bathula}}, \bibinfo
  {author} {\bibfnamefont {N.~K.}\ \bibnamefont {Singh}}, \bibinfo {author}
  {\bibfnamefont {S.}~\bibnamefont {Bishnoi}}, \bibinfo {author} {\bibfnamefont
  {S.}~\bibnamefont {Auluck}}, \bibinfo {author} {\bibfnamefont {A.~K.}\
  \bibnamefont {Srivastava}}, \ and\ \bibinfo {author} {\bibfnamefont
  {A.}~\bibnamefont {Dhar}},\ }\href {\doibase 10.1039/C5RA23742D} {\bibfield
  {journal} {\bibinfo  {journal} {RSC Adv.}\ }\textbf {\bibinfo {volume} {6}},\
  \bibinfo {pages} {11562} (\bibinfo {year} {2016})}\BibitemShut {NoStop}%
\bibitem [{\citenamefont {Shao}\ \emph {et~al.}(2016)\citenamefont {Shao},
  \citenamefont {Tan}, \citenamefont {Liu}, \citenamefont {Jiang},\ and\
  \citenamefont {Jiang}}]{Shao2016}%
  \BibitemOpen
  \bibfield  {author} {\bibinfo {author} {\bibfnamefont {H.}~\bibnamefont
  {Shao}}, \bibinfo {author} {\bibfnamefont {X.}~\bibnamefont {Tan}}, \bibinfo
  {author} {\bibfnamefont {G.-Q.}\ \bibnamefont {Liu}}, \bibinfo {author}
  {\bibfnamefont {J.}~\bibnamefont {Jiang}}, \ and\ \bibinfo {author}
  {\bibfnamefont {H.}~\bibnamefont {Jiang}},\ }\href {\doibase
  10.1038/srep21035} {\bibfield  {journal} {\bibinfo  {journal} {Sci. Rep.}\
  }\textbf {\bibinfo {volume} {6}},\ \bibinfo {pages} {21035} (\bibinfo {year}
  {2016})}\BibitemShut {NoStop}%
\bibitem [{\citenamefont {{Sun Lee}}\ \emph {et~al.}(2013)\citenamefont {{Sun
  Lee}}, \citenamefont {An}, \citenamefont {Jeong}, \citenamefont {Choi},
  \citenamefont {{Soo Lim}}, \citenamefont {Seo}, \citenamefont {Park},
  \citenamefont {Park},\ and\ \citenamefont {Park}}]{Lee2013}%
  \BibitemOpen
  \bibfield  {author} {\bibinfo {author} {\bibfnamefont {D.}~\bibnamefont {{Sun
  Lee}}}, \bibinfo {author} {\bibfnamefont {T.-H.}\ \bibnamefont {An}},
  \bibinfo {author} {\bibfnamefont {M.}~\bibnamefont {Jeong}}, \bibinfo
  {author} {\bibfnamefont {H.-S.}\ \bibnamefont {Choi}}, \bibinfo {author}
  {\bibfnamefont {Y.}~\bibnamefont {{Soo Lim}}}, \bibinfo {author}
  {\bibfnamefont {W.-S.}\ \bibnamefont {Seo}}, \bibinfo {author} {\bibfnamefont
  {C.-H.}\ \bibnamefont {Park}}, \bibinfo {author} {\bibfnamefont
  {C.}~\bibnamefont {Park}}, \ and\ \bibinfo {author} {\bibfnamefont {H.-H.}\
  \bibnamefont {Park}},\ }\href {\doibase 10.1063/1.4837475} {\bibfield
  {journal} {\bibinfo  {journal} {App. Phys. Lett.}\ }\textbf {\bibinfo
  {volume} {103}},\ \bibinfo {pages} {232110} (\bibinfo {year}
  {2013})}\BibitemShut {NoStop}%
\bibitem [{\citenamefont {Naguib}\ \emph {et~al.}(2014)\citenamefont {Naguib},
  \citenamefont {Mochalin}, \citenamefont {Barsoum},\ and\ \citenamefont
  {Gogotsi}}]{Naguib2014}%
  \BibitemOpen
  \bibfield  {author} {\bibinfo {author} {\bibfnamefont {M.}~\bibnamefont
  {Naguib}}, \bibinfo {author} {\bibfnamefont {V.~N.}\ \bibnamefont
  {Mochalin}}, \bibinfo {author} {\bibfnamefont {M.~W.}\ \bibnamefont
  {Barsoum}}, \ and\ \bibinfo {author} {\bibfnamefont {Y.}~\bibnamefont
  {Gogotsi}},\ }\href {\doibase 10.1002/adma.201304138} {\bibfield  {journal}
  {\bibinfo  {journal} {Adv. Mater.}\ }\textbf {\bibinfo {volume} {26}},\
  \bibinfo {pages} {992} (\bibinfo {year} {2014})}\BibitemShut {NoStop}%
\bibitem [{\citenamefont {Pandey}\ \emph {et~al.}(2015)\citenamefont {Pandey},
  \citenamefont {Vojvodic}, \citenamefont {Thygesen},\ and\ \citenamefont
  {Jacobsen}}]{Pandey2015a}%
  \BibitemOpen
  \bibfield  {author} {\bibinfo {author} {\bibfnamefont {M.}~\bibnamefont
  {Pandey}}, \bibinfo {author} {\bibfnamefont {A.}~\bibnamefont {Vojvodic}},
  \bibinfo {author} {\bibfnamefont {K.~S.}\ \bibnamefont {Thygesen}}, \ and\
  \bibinfo {author} {\bibfnamefont {K.~W.}\ \bibnamefont {Jacobsen}},\ }\href
  {\doibase 10.1021/acs.jpclett.5b00353} {\bibfield  {journal} {\bibinfo
  {journal} {J. Phys. Chem. Lett.}\ }\textbf {\bibinfo {volume} {6}},\ \bibinfo
  {pages} {1577} (\bibinfo {year} {2015})}\BibitemShut {NoStop}%
\bibitem [{\citenamefont {Zhang}\ \emph {et~al.}(2018)\citenamefont {Zhang},
  \citenamefont {Zhang}, \citenamefont {Yao}, \citenamefont {Chen},
  \citenamefont {Zhao},\ and\ \citenamefont {Zhou}}]{Zhang2018}%
  \BibitemOpen
  \bibfield  {author} {\bibinfo {author} {\bibfnamefont {X.}~\bibnamefont
  {Zhang}}, \bibinfo {author} {\bibfnamefont {Z.}~\bibnamefont {Zhang}},
  \bibinfo {author} {\bibfnamefont {S.}~\bibnamefont {Yao}}, \bibinfo {author}
  {\bibfnamefont {A.}~\bibnamefont {Chen}}, \bibinfo {author} {\bibfnamefont
  {X.}~\bibnamefont {Zhao}}, \ and\ \bibinfo {author} {\bibfnamefont
  {Z.}~\bibnamefont {Zhou}},\ }\href {\doibase 10.1038/s41524-018-0070-2}
  {\bibfield  {journal} {\bibinfo  {journal} {npj Comp. Mater.}\ }\textbf
  {\bibinfo {volume} {4}},\ \bibinfo {pages} {13} (\bibinfo {year}
  {2018})}\BibitemShut {NoStop}%
\bibitem [{\citenamefont {Toberer}\ \emph {et~al.}(2010)\citenamefont
  {Toberer}, \citenamefont {May}, \citenamefont {Melot}, \citenamefont
  {Flage-Larsen},\ and\ \citenamefont {Snyder}}]{Toberer2010a}%
  \BibitemOpen
  \bibfield  {author} {\bibinfo {author} {\bibfnamefont {E.~S.}\ \bibnamefont
  {Toberer}}, \bibinfo {author} {\bibfnamefont {A.~F.}\ \bibnamefont {May}},
  \bibinfo {author} {\bibfnamefont {B.~C.}\ \bibnamefont {Melot}}, \bibinfo
  {author} {\bibfnamefont {E.}~\bibnamefont {Flage-Larsen}}, \ and\ \bibinfo
  {author} {\bibfnamefont {G.~J.}\ \bibnamefont {Snyder}},\ }\href {\doibase
  10.1039/B914172C} {\bibfield  {journal} {\bibinfo  {journal} {Dalton Trans.}\
  }\textbf {\bibinfo {volume} {39}},\ \bibinfo {pages} {1046} (\bibinfo {year}
  {2010})}\BibitemShut {NoStop}%
\bibitem [{\citenamefont {Gascoin}\ \emph {et~al.}(2005)\citenamefont
  {Gascoin}, \citenamefont {Ottensmann}, \citenamefont {Stark}, \citenamefont
  {Ha{\"{i}}le},\ and\ \citenamefont {Snyder}}]{Gascoin2005}%
  \BibitemOpen
  \bibfield  {author} {\bibinfo {author} {\bibfnamefont {F.}~\bibnamefont
  {Gascoin}}, \bibinfo {author} {\bibfnamefont {S.}~\bibnamefont {Ottensmann}},
  \bibinfo {author} {\bibfnamefont {D.}~\bibnamefont {Stark}}, \bibinfo
  {author} {\bibfnamefont {S.~M.}\ \bibnamefont {Ha{\"{i}}le}}, \ and\ \bibinfo
  {author} {\bibfnamefont {G.~J.}\ \bibnamefont {Snyder}},\ }\href {\doibase
  10.1002/adfm.200500043} {\bibfield  {journal} {\bibinfo  {journal} {Adv.
  Func. Mater.}\ }\textbf {\bibinfo {volume} {15}},\ \bibinfo {pages} {1860}
  (\bibinfo {year} {2005})}\BibitemShut {NoStop}%
\bibitem [{\citenamefont {Zhang}\ \emph {et~al.}(2008)\citenamefont {Zhang},
  \citenamefont {Zhao}, \citenamefont {Grin}, \citenamefont {Wang},
  \citenamefont {Tang}, \citenamefont {Man}, \citenamefont {Chen},\ and\
  \citenamefont {Yang}}]{Zhang2008}%
  \BibitemOpen
  \bibfield  {author} {\bibinfo {author} {\bibfnamefont {H.}~\bibnamefont
  {Zhang}}, \bibinfo {author} {\bibfnamefont {J.-T.}\ \bibnamefont {Zhao}},
  \bibinfo {author} {\bibfnamefont {Y.}~\bibnamefont {Grin}}, \bibinfo {author}
  {\bibfnamefont {X.-J.}\ \bibnamefont {Wang}}, \bibinfo {author}
  {\bibfnamefont {M.-B.}\ \bibnamefont {Tang}}, \bibinfo {author}
  {\bibfnamefont {Z.-Y.}\ \bibnamefont {Man}}, \bibinfo {author} {\bibfnamefont
  {H.-H.}\ \bibnamefont {Chen}}, \ and\ \bibinfo {author} {\bibfnamefont
  {X.-X.}\ \bibnamefont {Yang}},\ }\href {\doibase 10.1063/1.3001608}
  {\bibfield  {journal} {\bibinfo  {journal} {J. Chem. Phys.}\ }\textbf
  {\bibinfo {volume} {129}},\ \bibinfo {pages} {164713} (\bibinfo {year}
  {2008})}\BibitemShut {NoStop}%
\bibitem [{\citenamefont {Sun}\ and\ \citenamefont {Singh}(2017)}]{Sun2017}%
  \BibitemOpen
  \bibfield  {author} {\bibinfo {author} {\bibfnamefont {J.}~\bibnamefont
  {Sun}}\ and\ \bibinfo {author} {\bibfnamefont {D.~J.}\ \bibnamefont
  {Singh}},\ }\href {\doibase 10.1039/C6TA11234J} {\bibfield  {journal}
  {\bibinfo  {journal} {J. Mater. Chem. A}\ }\textbf {\bibinfo {volume} {5}},\
  \bibinfo {pages} {8499} (\bibinfo {year} {2017})}\BibitemShut {NoStop}%
\bibitem [{\citenamefont {May}\ \emph {et~al.}(2012)\citenamefont {May},
  \citenamefont {McGuire}, \citenamefont {Ma}, \citenamefont {Delaire},
  \citenamefont {Huq},\ and\ \citenamefont {Custelcean}}]{May2012}%
  \BibitemOpen
  \bibfield  {author} {\bibinfo {author} {\bibfnamefont {A.~F.}\ \bibnamefont
  {May}}, \bibinfo {author} {\bibfnamefont {M.~A.}\ \bibnamefont {McGuire}},
  \bibinfo {author} {\bibfnamefont {J.}~\bibnamefont {Ma}}, \bibinfo {author}
  {\bibfnamefont {O.}~\bibnamefont {Delaire}}, \bibinfo {author} {\bibfnamefont
  {A.}~\bibnamefont {Huq}}, \ and\ \bibinfo {author} {\bibfnamefont
  {R.}~\bibnamefont {Custelcean}},\ }\href {\doibase 10.1063/1.3681817}
  {\bibfield  {journal} {\bibinfo  {journal} {J. App. Phys.}\ }\textbf
  {\bibinfo {volume} {111}},\ \bibinfo {pages} {033708} (\bibinfo {year}
  {2012})}\BibitemShut {NoStop}%
\bibitem [{\citenamefont {Labb{\'{e}}}\ and\ \citenamefont
  {Bok}(1987)}]{Labbe1987}%
  \BibitemOpen
  \bibfield  {author} {\bibinfo {author} {\bibfnamefont {J.}~\bibnamefont
  {Labb{\'{e}}}}\ and\ \bibinfo {author} {\bibfnamefont {J.}~\bibnamefont
  {Bok}},\ }\href {\doibase 10.1209/0295-5075/3/11/012} {\bibfield  {journal}
  {\bibinfo  {journal} {Europhys. Lett.}\ }\textbf {\bibinfo {volume} {3}},\
  \bibinfo {pages} {1225} (\bibinfo {year} {1987})}\BibitemShut {NoStop}%
\bibitem [{\citenamefont {Massidda}\ \emph {et~al.}(1987)\citenamefont
  {Massidda}, \citenamefont {Yu}, \citenamefont {Freeman},\ and\ \citenamefont
  {Koelling}}]{Massidda1987}%
  \BibitemOpen
  \bibfield  {author} {\bibinfo {author} {\bibfnamefont {S.}~\bibnamefont
  {Massidda}}, \bibinfo {author} {\bibfnamefont {J.}~\bibnamefont {Yu}},
  \bibinfo {author} {\bibfnamefont {A.}~\bibnamefont {Freeman}}, \ and\
  \bibinfo {author} {\bibfnamefont {D.}~\bibnamefont {Koelling}},\ }\href
  {\doibase 10.1016/0375-9601(87)90806-1} {\bibfield  {journal} {\bibinfo
  {journal} {Phys. Lett. A}\ }\textbf {\bibinfo {volume} {122}},\ \bibinfo
  {pages} {198} (\bibinfo {year} {1987})}\BibitemShut {NoStop}%
\bibitem [{\citenamefont {Anderson}(1987)}]{Anderson1987}%
  \BibitemOpen
  \bibfield  {author} {\bibinfo {author} {\bibfnamefont {P.~W.}\ \bibnamefont
  {Anderson}},\ }\href {\doibase 10.1126/science.235.4793.1196} {\bibfield
  {journal} {\bibinfo  {journal} {Science}\ }\textbf {\bibinfo {volume}
  {235}},\ \bibinfo {pages} {1196} (\bibinfo {year} {1987})}\BibitemShut
  {NoStop}%
\bibitem [{\citenamefont {Leggett}(1999)}]{Leggett1999}%
  \BibitemOpen
  \bibfield  {author} {\bibinfo {author} {\bibfnamefont {A.~J.}\ \bibnamefont
  {Leggett}},\ }\href {\doibase 10.1103/PhysRevLett.83.392} {\bibfield
  {journal} {\bibinfo  {journal} {Phys. Rev. Lett.}\ }\textbf {\bibinfo
  {volume} {83}},\ \bibinfo {pages} {392} (\bibinfo {year} {1999})}\BibitemShut
  {NoStop}%
\bibitem [{\citenamefont {Logvenov}\ \emph {et~al.}(2009)\citenamefont
  {Logvenov}, \citenamefont {Gozar},\ and\ \citenamefont
  {Bozovic}}]{Logvenov699}%
  \BibitemOpen
  \bibfield  {author} {\bibinfo {author} {\bibfnamefont {G.}~\bibnamefont
  {Logvenov}}, \bibinfo {author} {\bibfnamefont {A.}~\bibnamefont {Gozar}}, \
  and\ \bibinfo {author} {\bibfnamefont {I.}~\bibnamefont {Bozovic}},\ }\href
  {\doibase 10.1126/science.1178863} {\bibfield  {journal} {\bibinfo  {journal}
  {Science}\ }\textbf {\bibinfo {volume} {326}},\ \bibinfo {pages} {699}
  (\bibinfo {year} {2009})}\BibitemShut {NoStop}%
\bibitem [{\citenamefont {Lukatskaya}\ \emph {et~al.}(2013)\citenamefont
  {Lukatskaya}, \citenamefont {Mashtalir}, \citenamefont {Ren}, \citenamefont
  {Dall'Agnese}, \citenamefont {Rozier}, \citenamefont {Taberna}, \citenamefont
  {Naguib}, \citenamefont {Simon}, \citenamefont {Barsoum},\ and\ \citenamefont
  {Gogotsi}}]{Lukatskaya2013}%
  \BibitemOpen
  \bibfield  {author} {\bibinfo {author} {\bibfnamefont {M.~R.}\ \bibnamefont
  {Lukatskaya}}, \bibinfo {author} {\bibfnamefont {O.}~\bibnamefont
  {Mashtalir}}, \bibinfo {author} {\bibfnamefont {C.~E.}\ \bibnamefont {Ren}},
  \bibinfo {author} {\bibfnamefont {Y.}~\bibnamefont {Dall'Agnese}}, \bibinfo
  {author} {\bibfnamefont {P.}~\bibnamefont {Rozier}}, \bibinfo {author}
  {\bibfnamefont {P.~L.}\ \bibnamefont {Taberna}}, \bibinfo {author}
  {\bibfnamefont {M.}~\bibnamefont {Naguib}}, \bibinfo {author} {\bibfnamefont
  {P.}~\bibnamefont {Simon}}, \bibinfo {author} {\bibfnamefont {M.~W.}\
  \bibnamefont {Barsoum}}, \ and\ \bibinfo {author} {\bibfnamefont
  {Y.}~\bibnamefont {Gogotsi}},\ }\href {\doibase 10.1126/science.1241488}
  {\bibfield  {journal} {\bibinfo  {journal} {Science}\ }\textbf {\bibinfo
  {volume} {341}},\ \bibinfo {pages} {1502} (\bibinfo {year}
  {2013})}\BibitemShut {NoStop}%
\bibitem [{\citenamefont {Papoian}\ and\ \citenamefont
  {Hoffmann}(2000)}]{Papoian2000}%
  \BibitemOpen
  \bibfield  {author} {\bibinfo {author} {\bibfnamefont {G.~A.}\ \bibnamefont
  {Papoian}}\ and\ \bibinfo {author} {\bibfnamefont {R.}~\bibnamefont
  {Hoffmann}},\ }\href {\doibase
  10.1002/1521-3773(20000717)39:14<2408::AID-ANIE2408>3.0.CO;2-U} {\bibfield
  {journal} {\bibinfo  {journal} {Angewandte Chemie Inter. Ed.}\ }\textbf
  {\bibinfo {volume} {39}},\ \bibinfo {pages} {2408} (\bibinfo {year}
  {2000})}\BibitemShut {NoStop}%
\bibitem [{\citenamefont {Yan}\ \emph {et~al.}(2015)\citenamefont {Yan},
  \citenamefont {Gorai}, \citenamefont {Ortiz}, \citenamefont {Miller},
  \citenamefont {Barnett}, \citenamefont {Mason}, \citenamefont
  {Stevanovi{\'{c}}},\ and\ \citenamefont {Toberer}}]{Yan2015}%
  \BibitemOpen
  \bibfield  {author} {\bibinfo {author} {\bibfnamefont {J.}~\bibnamefont
  {Yan}}, \bibinfo {author} {\bibfnamefont {P.}~\bibnamefont {Gorai}}, \bibinfo
  {author} {\bibfnamefont {B.}~\bibnamefont {Ortiz}}, \bibinfo {author}
  {\bibfnamefont {S.}~\bibnamefont {Miller}}, \bibinfo {author} {\bibfnamefont
  {S.~A.}\ \bibnamefont {Barnett}}, \bibinfo {author} {\bibfnamefont
  {T.}~\bibnamefont {Mason}}, \bibinfo {author} {\bibfnamefont
  {V.}~\bibnamefont {Stevanovi{\'{c}}}}, \ and\ \bibinfo {author}
  {\bibfnamefont {E.~S.}\ \bibnamefont {Toberer}},\ }\href {\doibase
  10.1039/C4EE03157A} {\bibfield  {journal} {\bibinfo  {journal} {Energy
  Environ. Sci.}\ }\textbf {\bibinfo {volume} {8}},\ \bibinfo {pages} {983}
  (\bibinfo {year} {2015})}\BibitemShut {NoStop}%
\bibitem [{\citenamefont {Huang}\ \emph {et~al.}(2018)\citenamefont {Huang},
  \citenamefont {Liu}, \citenamefont {Fan}, \citenamefont {Jiang},
  \citenamefont {Liang}, \citenamefont {Cao}, \citenamefont {Liang},\ and\
  \citenamefont {Shi}}]{Huang2018}%
  \BibitemOpen
  \bibfield  {author} {\bibinfo {author} {\bibfnamefont {S.}~\bibnamefont
  {Huang}}, \bibinfo {author} {\bibfnamefont {H.~J.}\ \bibnamefont {Liu}},
  \bibinfo {author} {\bibfnamefont {D.~D.}\ \bibnamefont {Fan}}, \bibinfo
  {author} {\bibfnamefont {P.~H.}\ \bibnamefont {Jiang}}, \bibinfo {author}
  {\bibfnamefont {J.~H.}\ \bibnamefont {Liang}}, \bibinfo {author}
  {\bibfnamefont {G.~H.}\ \bibnamefont {Cao}}, \bibinfo {author} {\bibfnamefont
  {R.~Z.}\ \bibnamefont {Liang}}, \ and\ \bibinfo {author} {\bibfnamefont
  {J.}~\bibnamefont {Shi}},\ }\href {\doibase 10.1021/acs.jpcc.8b00099}
  {\bibfield  {journal} {\bibinfo  {journal} {The Journal of Physical Chemistry
  C}\ }\textbf {\bibinfo {volume} {122}},\ \bibinfo {pages} {4217} (\bibinfo
  {year} {2018})}\BibitemShut {NoStop}%
\bibitem [{\citenamefont {Ortiz}\ \emph {et~al.}(2017)\citenamefont {Ortiz},
  \citenamefont {Gorai}, \citenamefont {Krishna}, \citenamefont {Mow},
  \citenamefont {Lopez}, \citenamefont {McKinney}, \citenamefont
  {Stevanovi{\'{c}}},\ and\ \citenamefont {Toberer}}]{Ortiz2017}%
  \BibitemOpen
  \bibfield  {author} {\bibinfo {author} {\bibfnamefont {B.~R.}\ \bibnamefont
  {Ortiz}}, \bibinfo {author} {\bibfnamefont {P.}~\bibnamefont {Gorai}},
  \bibinfo {author} {\bibfnamefont {L.}~\bibnamefont {Krishna}}, \bibinfo
  {author} {\bibfnamefont {R.}~\bibnamefont {Mow}}, \bibinfo {author}
  {\bibfnamefont {A.}~\bibnamefont {Lopez}}, \bibinfo {author} {\bibfnamefont
  {R.}~\bibnamefont {McKinney}}, \bibinfo {author} {\bibfnamefont
  {V.}~\bibnamefont {Stevanovi{\'{c}}}}, \ and\ \bibinfo {author}
  {\bibfnamefont {E.~S.}\ \bibnamefont {Toberer}},\ }\href {\doibase
  10.1039/C6TA09532A} {\bibfield  {journal} {\bibinfo  {journal} {J. Mater.
  Chem. A}\ }\textbf {\bibinfo {volume} {5}},\ \bibinfo {pages} {4036}
  (\bibinfo {year} {2017})}\BibitemShut {NoStop}%
\bibitem [{\citenamefont {Belsky}\ \emph {et~al.}(2002)\citenamefont {Belsky},
  \citenamefont {Hellenbrandt}, \citenamefont {Karen},\ and\ \citenamefont
  {Luksch}}]{Belsky2002}%
  \BibitemOpen
  \bibfield  {author} {\bibinfo {author} {\bibfnamefont {A.}~\bibnamefont
  {Belsky}}, \bibinfo {author} {\bibfnamefont {M.}~\bibnamefont
  {Hellenbrandt}}, \bibinfo {author} {\bibfnamefont {V.~L.}\ \bibnamefont
  {Karen}}, \ and\ \bibinfo {author} {\bibfnamefont {P.}~\bibnamefont
  {Luksch}},\ }\href {\doibase 10.1107/S0108768102006948} {\bibfield  {journal}
  {\bibinfo  {journal} {Acta Crystallographica Sec. B Struct. Sci.}\ }\textbf
  {\bibinfo {volume} {58}},\ \bibinfo {pages} {364} (\bibinfo {year}
  {2002})}\BibitemShut {NoStop}%
\bibitem [{\citenamefont {Manna}\ \emph {et~al.}(2018)\citenamefont {Manna},
  \citenamefont {Gorai}, \citenamefont {Brennecka}, \citenamefont {Ciobanu},\
  and\ \citenamefont {Stevanovi{\'{c}}}}]{Manna2018}%
  \BibitemOpen
  \bibfield  {author} {\bibinfo {author} {\bibfnamefont {S.}~\bibnamefont
  {Manna}}, \bibinfo {author} {\bibfnamefont {P.}~\bibnamefont {Gorai}},
  \bibinfo {author} {\bibfnamefont {G.~L.}\ \bibnamefont {Brennecka}}, \bibinfo
  {author} {\bibfnamefont {C.~V.}\ \bibnamefont {Ciobanu}}, \ and\ \bibinfo
  {author} {\bibfnamefont {V.}~\bibnamefont {Stevanovi{\'{c}}}},\ }\href@noop
  {} {}\bibinfo {howpublished}
  {\href{https://arxiv.org/abs/1804.10997}{arXiv:1804.10997}} (\bibinfo {year}
  {2018})\BibitemShut {NoStop}%
\bibitem [{\citenamefont {Stevanovi{\'{c}}}\ \emph
  {et~al.}(2014{\natexlab{a}})\citenamefont {Stevanovi{\'{c}}}, \citenamefont
  {Lany}, \citenamefont {Ginley}, \citenamefont {Tumas},\ and\ \citenamefont
  {Zunger}}]{Stevanovic2014a}%
  \BibitemOpen
  \bibfield  {author} {\bibinfo {author} {\bibfnamefont {V.}~\bibnamefont
  {Stevanovi{\'{c}}}}, \bibinfo {author} {\bibfnamefont {S.}~\bibnamefont
  {Lany}}, \bibinfo {author} {\bibfnamefont {D.~S.}\ \bibnamefont {Ginley}},
  \bibinfo {author} {\bibfnamefont {W.}~\bibnamefont {Tumas}}, \ and\ \bibinfo
  {author} {\bibfnamefont {A.}~\bibnamefont {Zunger}},\ }\href {\doibase
  10.1039/c3cp54589j} {\bibfield  {journal} {\bibinfo  {journal} {Phys. Chem.
  Chem. Phys.}\ }\textbf {\bibinfo {volume} {16}},\ \bibinfo {pages} {3706}
  (\bibinfo {year} {2014}{\natexlab{a}})}\BibitemShut {NoStop}%
\bibitem [{\citenamefont {Stevanovi{\'{c}}}\ \emph
  {et~al.}(2014{\natexlab{b}})\citenamefont {Stevanovi{\'{c}}}, \citenamefont
  {Hartman}, \citenamefont {Jaramillo}, \citenamefont {Ramanathan},
  \citenamefont {Buonassisi},\ and\ \citenamefont {Graf}}]{Stevanovic2014b}%
  \BibitemOpen
  \bibfield  {author} {\bibinfo {author} {\bibfnamefont {V.}~\bibnamefont
  {Stevanovi{\'{c}}}}, \bibinfo {author} {\bibfnamefont {K.}~\bibnamefont
  {Hartman}}, \bibinfo {author} {\bibfnamefont {R.}~\bibnamefont {Jaramillo}},
  \bibinfo {author} {\bibfnamefont {S.}~\bibnamefont {Ramanathan}}, \bibinfo
  {author} {\bibfnamefont {T.}~\bibnamefont {Buonassisi}}, \ and\ \bibinfo
  {author} {\bibfnamefont {P.}~\bibnamefont {Graf}},\ }\href {\doibase
  10.1063/1.4879558} {\bibfield  {journal} {\bibinfo  {journal} {App. Phys.
  Lett.}\ }\textbf {\bibinfo {volume} {104}},\ \bibinfo {pages} {211603}
  (\bibinfo {year} {2014}{\natexlab{b}})}\BibitemShut {NoStop}%
\bibitem [{Note1()}]{Note1}%
  \BibitemOpen
  \bibinfo {note} {We avoid the $f$-electron materials because DFT is known to
  be highly inaccurate in describing the electron correlation in $f$-electron
  systems and also suffers from convergence issues due to the highly localized
  $f$-electron states.}\BibitemShut {Stop}%
\bibitem [{\citenamefont {Jiang}\ \emph {et~al.}(1998)\citenamefont {Jiang},
  \citenamefont {Lough}, \citenamefont {Ozin}, \citenamefont {Bedard},\ and\
  \citenamefont {Broach}}]{Jiang1998}%
  \BibitemOpen
  \bibfield  {author} {\bibinfo {author} {\bibfnamefont {T.}~\bibnamefont
  {Jiang}}, \bibinfo {author} {\bibfnamefont {A.}~\bibnamefont {Lough}},
  \bibinfo {author} {\bibfnamefont {G.~A.}\ \bibnamefont {Ozin}}, \bibinfo
  {author} {\bibfnamefont {R.~L.}\ \bibnamefont {Bedard}}, \ and\ \bibinfo
  {author} {\bibfnamefont {R.}~\bibnamefont {Broach}},\ }\href {\doibase
  10.1039/a706279f} {\bibfield  {journal} {\bibinfo  {journal} {J. Mater.
  Chem.}\ }\textbf {\bibinfo {volume} {8}},\ \bibinfo {pages} {721} (\bibinfo
  {year} {1998})}\BibitemShut {NoStop}%
\bibitem [{\citenamefont {Aydinol}\ \emph {et~al.}(1997)\citenamefont
  {Aydinol}, \citenamefont {Kohan}, \citenamefont {Ceder}, \citenamefont
  {Cho},\ and\ \citenamefont {Joannopoulos}}]{Aydinol1997}%
  \BibitemOpen
  \bibfield  {author} {\bibinfo {author} {\bibfnamefont {M.~K.}\ \bibnamefont
  {Aydinol}}, \bibinfo {author} {\bibfnamefont {A.~F.}\ \bibnamefont {Kohan}},
  \bibinfo {author} {\bibfnamefont {G.}~\bibnamefont {Ceder}}, \bibinfo
  {author} {\bibfnamefont {K.}~\bibnamefont {Cho}}, \ and\ \bibinfo {author}
  {\bibfnamefont {J.}~\bibnamefont {Joannopoulos}},\ }\href {\doibase
  10.1103/PhysRevB.56.1354} {\bibfield  {journal} {\bibinfo  {journal} {Phys.
  Rev. B}\ }\textbf {\bibinfo {volume} {56}},\ \bibinfo {pages} {1354}
  (\bibinfo {year} {1997})}\BibitemShut {NoStop}%
\bibitem [{\citenamefont {Born}(1940)}]{Born1940}%
  \BibitemOpen
  \bibfield  {author} {\bibinfo {author} {\bibfnamefont {M.}~\bibnamefont
  {Born}},\ }\href {\doibase 10.1017/S0305004100017138} {\bibfield  {journal}
  {\bibinfo  {journal} {Math. Proc. Cambridge Philos. Soc.}\ }\textbf {\bibinfo
  {volume} {36}},\ \bibinfo {pages} {160} (\bibinfo {year} {1940})}\BibitemShut
  {NoStop}%
\bibitem [{\citenamefont {Karki}\ \emph {et~al.}(1997)\citenamefont {Karki},
  \citenamefont {Ackland},\ and\ \citenamefont {Crain}}]{Karki1997}%
  \BibitemOpen
  \bibfield  {author} {\bibinfo {author} {\bibfnamefont {B.~B.}\ \bibnamefont
  {Karki}}, \bibinfo {author} {\bibfnamefont {G.~J.}\ \bibnamefont {Ackland}},
  \ and\ \bibinfo {author} {\bibfnamefont {J.}~\bibnamefont {Crain}},\ }\href
  {\doibase 10.1088/0953-8984/9/41/005} {\bibfield  {journal} {\bibinfo
  {journal} {J. Phys. Cond. Matt.}\ }\textbf {\bibinfo {volume} {9}},\ \bibinfo
  {pages} {8579} (\bibinfo {year} {1997})}\BibitemShut {NoStop}%
\bibitem [{\citenamefont {Morris}\ and\ \citenamefont
  {Krenn}(2000)}]{Morris2000}%
  \BibitemOpen
  \bibfield  {author} {\bibinfo {author} {\bibfnamefont {J.~W.}\ \bibnamefont
  {Morris}}\ and\ \bibinfo {author} {\bibfnamefont {C.~R.}\ \bibnamefont
  {Krenn}},\ }\href {\doibase 10.1080/01418610008223897} {\bibfield  {journal}
  {\bibinfo  {journal} {Philos. Mag. A}\ }\textbf {\bibinfo {volume} {80}},\
  \bibinfo {pages} {2827} (\bibinfo {year} {2000})}\BibitemShut {NoStop}%
\bibitem [{\citenamefont {Mouhat}\ and\ \citenamefont
  {Coudert}(2014)}]{Mouhat2014}%
  \BibitemOpen
  \bibfield  {author} {\bibinfo {author} {\bibfnamefont {F.}~\bibnamefont
  {Mouhat}}\ and\ \bibinfo {author} {\bibfnamefont {F.~X.}\ \bibnamefont
  {Coudert}},\ }\href {\doibase 10.1103/PhysRevB.90.224104} {\bibfield
  {journal} {\bibinfo  {journal} {Phys. Rev. B}\ }\textbf {\bibinfo {volume}
  {90}},\ \bibinfo {pages} {224104} (\bibinfo {year} {2014})}\BibitemShut
  {NoStop}%
\bibitem [{\citenamefont {de~Jong}\ \emph {et~al.}(2015)\citenamefont
  {de~Jong}, \citenamefont {Chen}, \citenamefont {Angsten}, \citenamefont
  {Jain}, \citenamefont {Notestine}, \citenamefont {Gamst}, \citenamefont
  {Sluiter}, \citenamefont {{Krishna Ande}}, \citenamefont {van~der Zwaag},
  \citenamefont {Plata}, \citenamefont {Toher}, \citenamefont {Curtarolo},
  \citenamefont {Ceder}, \citenamefont {Persson},\ and\ \citenamefont
  {Asta}}]{DeJong2015}%
  \BibitemOpen
  \bibfield  {author} {\bibinfo {author} {\bibfnamefont {M.}~\bibnamefont
  {de~Jong}}, \bibinfo {author} {\bibfnamefont {W.}~\bibnamefont {Chen}},
  \bibinfo {author} {\bibfnamefont {T.}~\bibnamefont {Angsten}}, \bibinfo
  {author} {\bibfnamefont {A.}~\bibnamefont {Jain}}, \bibinfo {author}
  {\bibfnamefont {R.}~\bibnamefont {Notestine}}, \bibinfo {author}
  {\bibfnamefont {A.}~\bibnamefont {Gamst}}, \bibinfo {author} {\bibfnamefont
  {M.}~\bibnamefont {Sluiter}}, \bibinfo {author} {\bibfnamefont
  {C.}~\bibnamefont {{Krishna Ande}}}, \bibinfo {author} {\bibfnamefont
  {S.}~\bibnamefont {van~der Zwaag}}, \bibinfo {author} {\bibfnamefont {J.~J.}\
  \bibnamefont {Plata}}, \bibinfo {author} {\bibfnamefont {C.}~\bibnamefont
  {Toher}}, \bibinfo {author} {\bibfnamefont {S.}~\bibnamefont {Curtarolo}},
  \bibinfo {author} {\bibfnamefont {G.}~\bibnamefont {Ceder}}, \bibinfo
  {author} {\bibfnamefont {K.~A.}\ \bibnamefont {Persson}}, \ and\ \bibinfo
  {author} {\bibfnamefont {M.}~\bibnamefont {Asta}},\ }\href {\doibase
  10.1038/sdata.2015.9} {\bibfield  {journal} {\bibinfo  {journal} {Sci. Data}\
  }\textbf {\bibinfo {volume} {2}},\ \bibinfo {pages} {150009} (\bibinfo {year}
  {2015})}\BibitemShut {NoStop}%
\bibitem [{\citenamefont {Hill}(1952)}]{Hill1952}%
  \BibitemOpen
  \bibfield  {author} {\bibinfo {author} {\bibfnamefont {R.}~\bibnamefont
  {Hill}},\ }\href {\doibase 10.1088/0370-1298/65/5/307} {\bibfield  {journal}
  {\bibinfo  {journal} {Proc. Phys. Soc. Sec. A}\ }\textbf {\bibinfo {volume}
  {65}},\ \bibinfo {pages} {349} (\bibinfo {year} {1952})}\BibitemShut
  {NoStop}%
\bibitem [{\citenamefont {Ranganathan}\ and\ \citenamefont
  {Ostoja-Starzewski}(2008)}]{Ranganathan2008}%
  \BibitemOpen
  \bibfield  {author} {\bibinfo {author} {\bibfnamefont {S.~I.}\ \bibnamefont
  {Ranganathan}}\ and\ \bibinfo {author} {\bibfnamefont {M.}~\bibnamefont
  {Ostoja-Starzewski}},\ }\href {\doibase 10.1103/PhysRevLett.101.055504}
  {\bibfield  {journal} {\bibinfo  {journal} {Physical Review Letters}\
  }\textbf {\bibinfo {volume} {101}},\ \bibinfo {pages} {055504} (\bibinfo
  {year} {2008})}\BibitemShut {NoStop}%
\bibitem [{\citenamefont {Peng}\ \emph {et~al.}(2015)\citenamefont {Peng},
  \citenamefont {Duan},\ and\ \citenamefont {Sun}}]{Peng2015}%
  \BibitemOpen
  \bibfield  {author} {\bibinfo {author} {\bibfnamefont {M.}~\bibnamefont
  {Peng}}, \bibinfo {author} {\bibfnamefont {Y.}~\bibnamefont {Duan}}, \ and\
  \bibinfo {author} {\bibfnamefont {Y.}~\bibnamefont {Sun}},\ }\href {\doibase
  10.1016/j.commatsci.2014.10.060} {\bibfield  {journal} {\bibinfo  {journal}
  {Comp. Mater. Sci.}\ }\textbf {\bibinfo {volume} {98}},\ \bibinfo {pages}
  {311} (\bibinfo {year} {2015})}\BibitemShut {NoStop}%
\bibitem [{\citenamefont {Kube}(2016)}]{Kube2016a}%
  \BibitemOpen
  \bibfield  {author} {\bibinfo {author} {\bibfnamefont {C.~M.}\ \bibnamefont
  {Kube}},\ }\href {\doibase 10.1063/1.4962996} {\bibfield  {journal} {\bibinfo
   {journal} {AIP Advances}\ }\textbf {\bibinfo {volume} {6}},\ \bibinfo
  {pages} {095209} (\bibinfo {year} {2016})}\BibitemShut {NoStop}%
\bibitem [{\citenamefont {Ting}(2006)}]{Ting2005}%
  \BibitemOpen
  \bibfield  {author} {\bibinfo {author} {\bibfnamefont {T.~C.~T.}\
  \bibnamefont {Ting}},\ }\href {\doibase 10.1007/s10659-005-9016-2} {\bibfield
   {journal} {\bibinfo  {journal} {J. Elasticity}\ }\textbf {\bibinfo {volume}
  {81}},\ \bibinfo {pages} {271} (\bibinfo {year} {2006})}\BibitemShut
  {NoStop}%
\bibitem [{\citenamefont {Kresse}\ and\ \citenamefont
  {Furthm{\"{u}}ller}(1996{\natexlab{a}})}]{Kresse1996}%
  \BibitemOpen
  \bibfield  {author} {\bibinfo {author} {\bibfnamefont {G.}~\bibnamefont
  {Kresse}}\ and\ \bibinfo {author} {\bibfnamefont {J.}~\bibnamefont
  {Furthm{\"{u}}ller}},\ }\href {\doibase 10.1016/0927-0256(96)00008-0}
  {\bibfield  {journal} {\bibinfo  {journal} {Comp. Mater. Sci.}\ }\textbf
  {\bibinfo {volume} {6}},\ \bibinfo {pages} {15} (\bibinfo {year}
  {1996}{\natexlab{a}})}\BibitemShut {NoStop}%
\bibitem [{\citenamefont {Kresse}\ and\ \citenamefont
  {Furthm{\"{u}}ller}(1996{\natexlab{b}})}]{Kresse1996a}%
  \BibitemOpen
  \bibfield  {author} {\bibinfo {author} {\bibfnamefont {G.}~\bibnamefont
  {Kresse}}\ and\ \bibinfo {author} {\bibfnamefont {J.}~\bibnamefont
  {Furthm{\"{u}}ller}},\ }\href {\doibase 10.1103/PhysRevB.54.11169} {\bibfield
   {journal} {\bibinfo  {journal} {Phys. Rev. B}\ }\textbf {\bibinfo {volume}
  {54}},\ \bibinfo {pages} {11169} (\bibinfo {year}
  {1996}{\natexlab{b}})}\BibitemShut {NoStop}%
\bibitem [{\citenamefont {Bl{\"{o}}chl}(1994)}]{Blochl1994}%
  \BibitemOpen
  \bibfield  {author} {\bibinfo {author} {\bibfnamefont {P.~E.}\ \bibnamefont
  {Bl{\"{o}}chl}},\ }\href {\doibase 10.1103/PhysRevB.50.17953} {\bibfield
  {journal} {\bibinfo  {journal} {Phys. Rev. B}\ }\textbf {\bibinfo {volume}
  {50}},\ \bibinfo {pages} {17953} (\bibinfo {year} {1994})}\BibitemShut
  {NoStop}%
\bibitem [{\citenamefont {Kresse}\ and\ \citenamefont
  {Joubert}(1999)}]{Kresse1999}%
  \BibitemOpen
  \bibfield  {author} {\bibinfo {author} {\bibfnamefont {G.}~\bibnamefont
  {Kresse}}\ and\ \bibinfo {author} {\bibfnamefont {D.}~\bibnamefont
  {Joubert}},\ }\href {\doibase 10.1103/PhysRevB.59.1758} {\bibfield  {journal}
  {\bibinfo  {journal} {Phys. Rev. B}\ }\textbf {\bibinfo {volume} {59}},\
  \bibinfo {pages} {1758} (\bibinfo {year} {1999})}\BibitemShut {NoStop}%
\bibitem [{\citenamefont {Perdew}\ \emph {et~al.}(1996)\citenamefont {Perdew},
  \citenamefont {Burke},\ and\ \citenamefont {Ernzerhof}}]{Perdew1996}%
  \BibitemOpen
  \bibfield  {author} {\bibinfo {author} {\bibfnamefont {J.~P.}\ \bibnamefont
  {Perdew}}, \bibinfo {author} {\bibfnamefont {K.}~\bibnamefont {Burke}}, \
  and\ \bibinfo {author} {\bibfnamefont {M.}~\bibnamefont {Ernzerhof}},\ }\href
  {\doibase 10.1103/PhysRevLett.77.3865} {\bibfield  {journal} {\bibinfo
  {journal} {Phys. Rev. Lett.}\ }\textbf {\bibinfo {volume} {77}},\ \bibinfo
  {pages} {3865} (\bibinfo {year} {1996})}\BibitemShut {NoStop}%
\bibitem [{\citenamefont {Monkhorst}\ and\ \citenamefont
  {Pack}(1976)}]{Monkhorst1976}%
  \BibitemOpen
  \bibfield  {author} {\bibinfo {author} {\bibfnamefont {H.~J.}\ \bibnamefont
  {Monkhorst}}\ and\ \bibinfo {author} {\bibfnamefont {J.~D.}\ \bibnamefont
  {Pack}},\ }\href {\doibase 10.1103/PhysRevB.13.5188} {\bibfield  {journal}
  {\bibinfo  {journal} {Phys. Rev. B}\ }\textbf {\bibinfo {volume} {13}},\
  \bibinfo {pages} {5188} (\bibinfo {year} {1976})}\BibitemShut {NoStop}%
\bibitem [{\citenamefont {Klime{\v{s}}}\ \emph {et~al.}(2011)\citenamefont
  {Klime{\v{s}}}, \citenamefont {Bowler},\ and\ \citenamefont
  {Michaelides}}]{Klimes2011}%
  \BibitemOpen
  \bibfield  {author} {\bibinfo {author} {\bibfnamefont {J.}~\bibnamefont
  {Klime{\v{s}}}}, \bibinfo {author} {\bibfnamefont {D.~R.}\ \bibnamefont
  {Bowler}}, \ and\ \bibinfo {author} {\bibfnamefont {A.}~\bibnamefont
  {Michaelides}},\ }\href {\doibase 10.1103/PhysRevB.83.195131} {\bibfield
  {journal} {\bibinfo  {journal} {Phys. Rev. B}\ }\textbf {\bibinfo {volume}
  {83}},\ \bibinfo {pages} {195131} (\bibinfo {year} {2011})}\BibitemShut
  {NoStop}%
\bibitem [{\citenamefont {{Le Page}}\ and\ \citenamefont
  {Saxe}(2002)}]{LePage2002}%
  \BibitemOpen
  \bibfield  {author} {\bibinfo {author} {\bibfnamefont {Y.}~\bibnamefont {{Le
  Page}}}\ and\ \bibinfo {author} {\bibfnamefont {P.}~\bibnamefont {Saxe}},\
  }\href {\doibase 10.1103/PhysRevB.65.104104} {\bibfield  {journal} {\bibinfo
  {journal} {Phys. Rev. B}\ }\textbf {\bibinfo {volume} {65}},\ \bibinfo
  {pages} {104104} (\bibinfo {year} {2002})}\BibitemShut {NoStop}%
\bibitem [{\citenamefont {Wu}\ \emph {et~al.}(2005)\citenamefont {Wu},
  \citenamefont {Vanderbilt},\ and\ \citenamefont {Hamann}}]{Wu2005}%
  \BibitemOpen
  \bibfield  {author} {\bibinfo {author} {\bibfnamefont {X.}~\bibnamefont
  {Wu}}, \bibinfo {author} {\bibfnamefont {D.}~\bibnamefont {Vanderbilt}}, \
  and\ \bibinfo {author} {\bibfnamefont {D.~R.}\ \bibnamefont {Hamann}},\
  }\href {\doibase 10.1103/PhysRevB.72.035105} {\bibfield  {journal} {\bibinfo
  {journal} {Phys. Rev. B}\ }\textbf {\bibinfo {volume} {72}},\ \bibinfo
  {pages} {035105} (\bibinfo {year} {2005})}\BibitemShut {NoStop}%
\bibitem [{Pyl()}]{Pylada}%
  \BibitemOpen
  \href@noop {} {\enquote {\bibinfo {title} {Pylada: A python framework for
  high-throughput first-principles calculations},}\ }\bibinfo {howpublished}
  {\href{https://github.com/pylada}{github.com/pylada}}\BibitemShut {NoStop}%
\bibitem [{\citenamefont {Corbett}(2010)}]{Corbett2010}%
  \BibitemOpen
  \bibfield  {author} {\bibinfo {author} {\bibfnamefont {J.~D.}\ \bibnamefont
  {Corbett}},\ }\href {\doibase 10.1021/ic901305g} {\bibfield  {journal}
  {\bibinfo  {journal} {Inorg. Chem.}\ }\textbf {\bibinfo {volume} {49}},\
  \bibinfo {pages} {13} (\bibinfo {year} {2010})}\BibitemShut {NoStop}%
\bibitem [{\citenamefont {Eisenmann}\ \emph {et~al.}(1972)\citenamefont
  {Eisenmann}, \citenamefont {May}, \citenamefont {M{\"{u}}ller},\ and\
  \citenamefont {Sch{\"{a}}fer}}]{Eisenmann1972}%
  \BibitemOpen
  \bibfield  {author} {\bibinfo {author} {\bibfnamefont {B.}~\bibnamefont
  {Eisenmann}}, \bibinfo {author} {\bibfnamefont {N.}~\bibnamefont {May}},
  \bibinfo {author} {\bibfnamefont {W.}~\bibnamefont {M{\"{u}}ller}}, \ and\
  \bibinfo {author} {\bibfnamefont {H.}~\bibnamefont {Sch{\"{a}}fer}},\ }\href
  {\doibase 10.1515/znb-1972-1008} {\bibfield  {journal} {\bibinfo  {journal}
  {Zeitschrift f{\"{u}}r Naturforschung B}\ }\textbf {\bibinfo {volume} {27}},\
  \bibinfo {pages} {1155} (\bibinfo {year} {1972})}\BibitemShut {NoStop}%
\bibitem [{\citenamefont {Barsoum}(2000)}]{Barsoum2000}%
  \BibitemOpen
  \bibfield  {author} {\bibinfo {author} {\bibfnamefont {M.~W.}\ \bibnamefont
  {Barsoum}},\ }\href {\doibase 10.1016/S0079-6786(00)00006-6} {\bibfield
  {journal} {\bibinfo  {journal} {Prog. Sol. State Chem.}\ }\textbf {\bibinfo
  {volume} {28}},\ \bibinfo {pages} {201} (\bibinfo {year} {2000})}\BibitemShut
  {NoStop}%
\bibitem [{\citenamefont {Zhong}\ \emph {et~al.}(1995)\citenamefont {Zhong},
  \citenamefont {Vanderbilt},\ and\ \citenamefont {Rabe}}]{Zhong1995}%
  \BibitemOpen
  \bibfield  {author} {\bibinfo {author} {\bibfnamefont {W.}~\bibnamefont
  {Zhong}}, \bibinfo {author} {\bibfnamefont {D.}~\bibnamefont {Vanderbilt}}, \
  and\ \bibinfo {author} {\bibfnamefont {K.~M.}\ \bibnamefont {Rabe}},\ }\href
  {\doibase 10.1103/PhysRevB.52.6301} {\bibfield  {journal} {\bibinfo
  {journal} {Phys. Rev. B}\ }\textbf {\bibinfo {volume} {52}},\ \bibinfo
  {pages} {6301} (\bibinfo {year} {1995})}\BibitemShut {NoStop}%
\bibitem [{\citenamefont {Fompeyrine}\ \emph {et~al.}(1999)\citenamefont
  {Fompeyrine}, \citenamefont {Seo},\ and\ \citenamefont
  {Locquet}}]{Fompeyrine1999}%
  \BibitemOpen
  \bibfield  {author} {\bibinfo {author} {\bibfnamefont {J.}~\bibnamefont
  {Fompeyrine}}, \bibinfo {author} {\bibfnamefont {J.}~\bibnamefont {Seo}}, \
  and\ \bibinfo {author} {\bibfnamefont {J.-P.}\ \bibnamefont {Locquet}},\
  }\href {\doibase 10.1016/S0955-2219(98)00463-4} {\bibfield  {journal}
  {\bibinfo  {journal} {J. Euro. Ceramic Soc.}\ }\textbf {\bibinfo {volume}
  {19}},\ \bibinfo {pages} {1493} (\bibinfo {year} {1999})}\BibitemShut
  {NoStop}%
\end{thebibliography}%
\end{document}